\documentclass[aps, prd, twocolumn, superscriptaddress,preprintnumbers,nofootinbib]{revtex4}

\usepackage{graphicx}
\usepackage{dcolumn}
\usepackage{amssymb}

\usepackage{color}

\def\ep{\epsilon}

\def\ka{\kappa}

\def\si{\sigma}

\newcommand{\ben}{\begin{equation}}
\newcommand{\een}{\end{equation}}
\newcommand{\bea}{\begin{eqnarray}}
\newcommand{\eea}{\end{eqnarray}}
\newcommand{\ba}{\begin{array}}
\newcommand{\ea}{\end{array}}
\newcommand{\bit}{\begin{itemize}}
\newcommand{\eit}{\end{itemize}}
\newcommand{\gmu}{$G\mu$}

\begin{document}

\newcommand{\fd}{f_{10}}
\newcommand{\As}{A_{\mathrm{s}}}
\newcommand{\At}{A_{\mathrm{t}}}
\newcommand{\ns}{n_{\mathrm{s}}}
\newcommand{\nt}{n_{\mathrm{t}}}
\newcommand{\Obhh}{\Omega_{\mathrm{b}}h^{2}}
\newcommand{\Omhh}{\Omega_{\mathrm{m}}h^{2}}
\newcommand{\Ol}{\Omega_{\Lambda}}

\newcommand{\CMB}{\textsc{CMB }}
\newcommand{\CMBEASY}{\textsc{CMBeasy }}
\newcommand{\TT}{\textsc{tt }}
\newcommand{\TE}{\textsc{te }}
\newcommand{\EE}{\textsc{ee }}
\newcommand{\BB}{\textsc{bb }}
\newcommand{\WMAP}{\textsc{WMAP }}
\newcommand{\UETC}{\textsc{uetc }}
\newcommand{\UETCs}{\textsc{uetc}s}
\newcommand{\ETC}{\textsc{etc }}
\newcommand{\ETCs}{\textsc{etc}s }
\newcommand{\HK}{\textsc{HKP }}
\newcommand{\BBN}{\textsc{BBN }}
\newcommand{\mcmc}{\textsc{MCMC }}

\newcommand{\PL}{$\mathcal{PL}$}
\newcommand{\PLr}{$\mathcal{PL}+r$}
\newcommand{\PLgmu}{$\mathcal{PL}+G\mu$}
\newcommand{\PLAH}{$\mathcal{PL}+G\mu[\mathrm{AH}]$}
\newcommand{\PLSL}{$\mathcal{PL}+G\mu[\mathrm{SL}]$}
\newcommand{\PLTX}{$\mathcal{PL}+G\mu[\mathrm{TX}]$}
\newcommand{\PLrgmu}{$\mathcal{PL}+r+G\mu$}
\newcommand{\PLralpha}{$\mathcal{PL}+r+\alpha_{\rm s}$}
\newcommand{\PLrgmualpha}{$\mathcal{PL}+r+G\mu+\alpha_{\rm s}$}
\newcommand{\PLdust}{$\mathcal{PL}+A_{\rm dust}$}
\newcommand{\PLgmudust}{$\mathcal{PL}+G\mu+A_{\rm dust}$}
\newcommand{\PLAHgmudust}{$\mathcal{PL}+G\mu[\mathrm{AH}]+A_{\rm dust}$}
\newcommand{\PLrdust}{$\mathcal{PL}+r+A_{\rm dust}$}
\newcommand{\PLrgmudust}{$\mathcal{PL}+r+G\mu+A_{\rm dust}$}

\newcommand{\half}{\frac{1}{2}}

\newcommand{\clover}{\textsc{C}$$\ell$$\textsc{over}}

\renewcommand{\d}{{\partial}}

\newcommand{\BHKU}{BHKU}
\newcommand{\Ob}{\ensuremath{\Omega_{\mathrm b}}}
\newcommand{\Oc}{\ensuremath{\Omega_{\mathrm c}}}
\newcommand{\Omm}{\ensuremath{\Omega_{\mathrm m}}}
\newcommand{\Ochh}{\ensuremath{\Oc h^{2}}}
\newcommand{\optdepth}{\tau}
\newcommand{\Asz}{A_\mathrm{SZ}}
\newcommand{\Ap}{A_\mathrm{p}}
\newcommand{\Cl}{\mathcal{C}_{$\ell$}}
\newcommand{\Dl}{\mathcal{D}_{$\ell$}}
\newcommand{\VEV}{\PHI_{0}}
\newcommand{\PHI}{\phi}
\newcommand{\conj}{^*}
\newcommand{\FT}[1]{\tilde{#1}}
\newcommand{\vect}[1]{\mathbf{#1}}
\newcommand{\diff}{\mathrm{d}}
\newcommand{\Sr}{^{\mathrm{S}}}
\newcommand{\Vr}[1]{\!\!\stackrel{\scriptstyle{\mathrm{V}}}{_{\!\!#1}}}
\newcommand{\Tr}[1]{\!\!\stackrel{\scriptstyle{\mathrm{T}}}{_{\!#1}}}
\newcommand{\MM}{\mathcal{M}}

\newcommand{\unit}[1]{\;\mathrm{#1}}
\newcommand{\Eq}[1]{Eq. (\ref{eqn:#1})}
\newcommand{\Eqnb}[1]{Eq. \ref{eqn:#1}}
\newcommand{\Fig}[1]{Fig. \ref{fig:#1}}
\newcommand{\Sec}[1]{Sec. \ref{sec:#1}}
\newcommand{\Table}[1]{Table \ref{tab:#1}}

\newcommand{\PLSZ}{PL$_{\rm SZ}$}
\newcommand{\HZSZ}{HZ$_{\rm SZ}$}

\newcommand{\mbh}[1]{\textcolor{blue}{MBH: #1}}
\newcommand{\ju}[1]{\textcolor{blue}{JU: #1}}

\newcommand{\Planck}{{\it Planck}}
\newcommand{\LCDM}{$\Lambda$CDM}

\title{Constraining topological defects with temperature and polarization anisotropies} 

\newcommand{\addressSussex}{Department of Physics \&
Astronomy, University of Sussex, Brighton, BN1 9QH, United Kingdom}
\newcommand{\addressGeneva}{D\'epartement de Physique Th\'eorique \& Center for Astroparticle Physics,
Universit\'e de Gen\`eve, Quai E.\ Ansermet 24, CH-1211 Gen\`eve 4, Switzerland}
\newcommand{\addressBilbao}{Department of Theoretical Physics, University of the Basque Country UPV/EHU,
48080 Bilbao, Spain}
\newcommand{\addressEdinburgh}{Institute for Astronomy, University of Edinburgh, Royal Observatory, Edinburgh EH9 3HJ,
United Kingdom}
\newcommand{\addressHelsinki}
{Department of Physics and Helsinki Institute of Physics, PL 64, FI-00014 University of Helsinki, Finland}
\newcommand{\addressAIMS}
{African Institute for Mathematical Sciences, 6 Melrose Road, Muizenberg, 7945, South Africa}

\author{Joanes Lizarraga}
\affiliation{\addressBilbao}

\author{Jon Urrestilla}
\affiliation{\addressBilbao}

\author{David Daverio}
\affiliation{\addressGeneva}

\author{Mark Hindmarsh}
\affiliation{\addressSussex}
\affiliation{\addressHelsinki}

\author{Martin Kunz}
\affiliation{\addressGeneva}
\affiliation{\addressAIMS}

\author{Andrew R.~Liddle}
\affiliation{\addressEdinburgh}

\date{\today}

\begin{abstract}
We analyse the possible contribution of topological defects to cosmic microwave anisotropies, both temperature and polarisation. 
We allow for the presence of both inflationary scalars and tensors, and of polarised dust foregrounds that may contribute to or dominate the B-mode polarisation signal. We confirm and quantify our previous statements that topological defects on their own are a poor fit to the B-mode signal.  
However, adding topological defects to a models with a tensor component or a dust component improves the fit around $\ell=200$. 
Fitting simultaneously to both temperature and polarisation data, we find that textures fit almost as well as tensors ($\Delta\chi^2 = 2.0$), while Abelian Higgs strings are ruled out as the sole source of the B-mode signal at low $\ell$. 
The 95\% confidence upper  limits on models combining defects and dust are
$G\mu < 2.7\times 10^{-7}$ (Abelian Higgs strings), 
$G\mu < 9.8\times 10^{-7}$ (semilocal strings) and 
$G\mu < 7.3\times 10^{-7}$ (textures), a small reduction on the {\em Planck} bounds.
The most economical fit overall is obtained by the standard $\Lambda$CDM model with a polarised dust component.
\end{abstract}

\maketitle

\section{Introduction}

The recent detection of B-mode polarisation \cite{Ade:2013gez,Ade:2014afa,Ade:2014xna} opens a new avenue for constraining models of the early Universe at very high energy. The claim by the BICEP2 collaboration that the B-mode polarisation on large angular scales \cite{Ade:2014xna} is caused by primordial inflationary tensor modes has generated great excitement, and stimulated the search for other possible B-mode polarization sources signalling new physics, such as cosmic defects \cite{Lizarraga:2014eaa,Moss:2014cra}, self-ordering scalar fields \cite{Durrer:2014raa} or primordial magnetic fields \cite{Bonvin:2014xia}. 
However, it has subsequently become apparent that conventional astrophysics can plausibly account for the entire observed B-mode signal with polarised dust emission \cite{Mortonson:2014bja,Flauger:2014qra}. 
At the very least it is clear that dust contamination must be explored alongside any proposed primordial contribution.

In a previous paper \cite{Lizarraga:2014eaa} we showed that the predicted spectra from defects had the wrong shape to entirely explain the observed B-mode signal at low multipoles, although a good fit could be obtained in combination with inflationary tensors. However, we did not consider the possibility of foreground contributions to the polarisation, nor did we analyse the BICEP2 data in combination with other cosmic microwave background (CMB) datasets.   In this article we complement our previous paper by providing a comprehensive analysis of the defect contribution to the microwave anisotropies, both in temperature and polarisation, allowing all three of the above signal sources, {i.e.}, inflationary gravitational waves, dust, and cosmic defects.

Cosmic defects produce B-mode polarization through both tensor and vector modes (see e.g.\ Refs.~\cite{VilShe94,Hindmarsh:1994re,Durrer:2001cg,Copeland:2009ga,Hindmarsh:2011qj,Figueroa:2012kw} for reviews). The relative proportions of scalar, vector and tensor perturbations are essentially fixed for a given type of defect, so a constraint on one of the modes will imply constraints on the others. It is worth noting that even though defects are highly constrained via the CMB temperature anisotropies \cite{Wyman:2005tu,Bevis:2007gh,Battye:2010xz,Dunkley:2010ge,Urrestilla:2011gr,Avgoustidis:2011ax,Ade:2013xla}, they can still contribute importantly to the B-mode polarization.

In our analysis we study three types of cosmic defects: Abelian Higgs strings \cite{Nielsen:1973cs}, O(4) global textures \cite{Turok:1990gw}, and semilocal strings \cite{Vachaspati:1991dz,Hindmarsh:1991jq,Achucarro:1999it,Hindmarsh:1992yy}. Other defect models exist, such as self-ordering scalar fields, global monopoles, and global strings. 
However, with the three types of defects under consideration we are able to obtain a global view of the interplay between cosmic defects and the other signals, and can also study where the differences between the different defect predictions are important.  The imprints of defects on the temperature and polarization power spectra are qualitatively similar 
\cite{Bevis:2006mj,Bevis:2007qz,Bevis:2010gj,Pen:1997ae,Durrer:1998rw,Durrer:2001cg,Urrestilla:2007sf,Urrestilla:2008jv},
though there are important quantitative differences.  

In the next section we describe the defect models we have considered. In Section~\ref{models} we describe the methodology, cosmological models and datasets that we have used. The results of the analysis are reported in Section~\ref{results}, and in the final section we present conclusions and discussion.

\section{CMB spectra from defects}
\label{defects}

The cosmic defects that are most strongly motivated by particle physics, as they arise from spontaneously-broken gauge symmetries, are 
cosmic strings \cite{VilShe94,Hindmarsh:1994re,Copeland:2011dx,Hindmarsh:2011qj}. They are predicted to form in many high-energy inflationary models \cite{Yokoyama:1988zza,Copeland:1994vg,Lyth:1998xn,Jeannerot:2003qv,Majumdar:2002hy,Sarangi:2002yt,Copeland:2003bj,Dvali:2003zj,Tye:2014tja}. 
Our two other examples of cosmic defects, textures and semilocal strings, arise when spontaneously-broken global symmetries are present.
 
The perturbation power spectra of topological defects can be calculated from numerical simulations of an underlying field theory in an expanding cosmological model \cite{Pen:1997ae,Durrer:1998rw,Bevis:2006mj,Urrestilla:2007sf}. 
Cosmic string spectra have been calculated for the Abelian Higgs model \cite{Bevis:2006mj,Bevis:2007qz,Bevis:2010gj}, 
textures in an O(4) non-linear \cite{Pen:1997ae,Durrer:1998rw,Durrer:2001cg} or linear \cite{Urrestilla:2007sf} $\si$-model,
and semilocal strings in a U(1) theory with an extra SU(2) global symmetry \cite{Urrestilla:2007sf}.

Another approach to model cosmic string networks, including those arising from superstring models, is based on simulating directly the evolution of  string-like objects based on the Nambu-Goto action  \cite{Albrecht:1984xv,Bennett:1987vf,Albrecht:1989mk,Bennett:1989ak,Bennett:1989yp,Allen:1990tv,Sakellariadou:1990nd,Vincent:1996rb}. There has been very intense work into understanding the loop generation and dynamics in this model \cite{Vanchurin:2005yb,Vanchurin:2005pa,Ringeval:2005kr,Olum:2006ix,Lorenz:2010sm,BlancoPillado:2010sy,BlancoPillado:2011dq}.  However, there is no numerical simulation of the Nambu-Goto model  calculating the full CMB temperature or polarization spectra, although there has been some work in that direction \cite{Landriau:2010cb,Ringeval:2010ca}. An alternative to full Nambu-Goto type simulations is afforded by the unconnected segment model (USM)   \cite{Albrecht:1997mz,Avgoustidis:2012gb,Pogosian:1999np}, which introduces an extra layer of modeling and can be tuned to mimic not only Nambu-Goto strings but also the behaviour of Abelian-Higgs string networks, in which case it gives a good approximation to the power spectra of the CMB anisotropies   \cite{Battye:2010xz}. 

There are also other approaches for the other defects considered in this work.
For textures, there is an analytic approximation in the  large $N$ limit of the O($N$) non-linear $\si$-model \cite{Kunz:1996ka}. In Ref.~\cite{Urrestilla:2007sf} a comparison between the linear and non-linear $\sigma$-model can be found, showing that they are very close. There is also a model describing the evolution of semilocal strings \cite{Nunes:2011sf,Achucarro:2013mga}. 

The defect spectra used in this paper were calculated in Refs.~\cite{Urrestilla:2007sf,Bevis:2010gj} using  a modified version of \CMBEASY \cite{Doran:2003sy}, with the best-fit parameters of the WMAP 7-year analysis \cite{Komatsu:2010fb}. We do not vary the cosmological model used for computing the defect spectra, as the spectra change little for the allowed range of cosmological parameters. Since the defect contribution is sub-dominant in the temperature power spectrum, the resulting inaccuracies in the parameter posteriors are insignificant.

\begin{figure}[b]
\resizebox{\columnwidth}{!}{\includegraphics{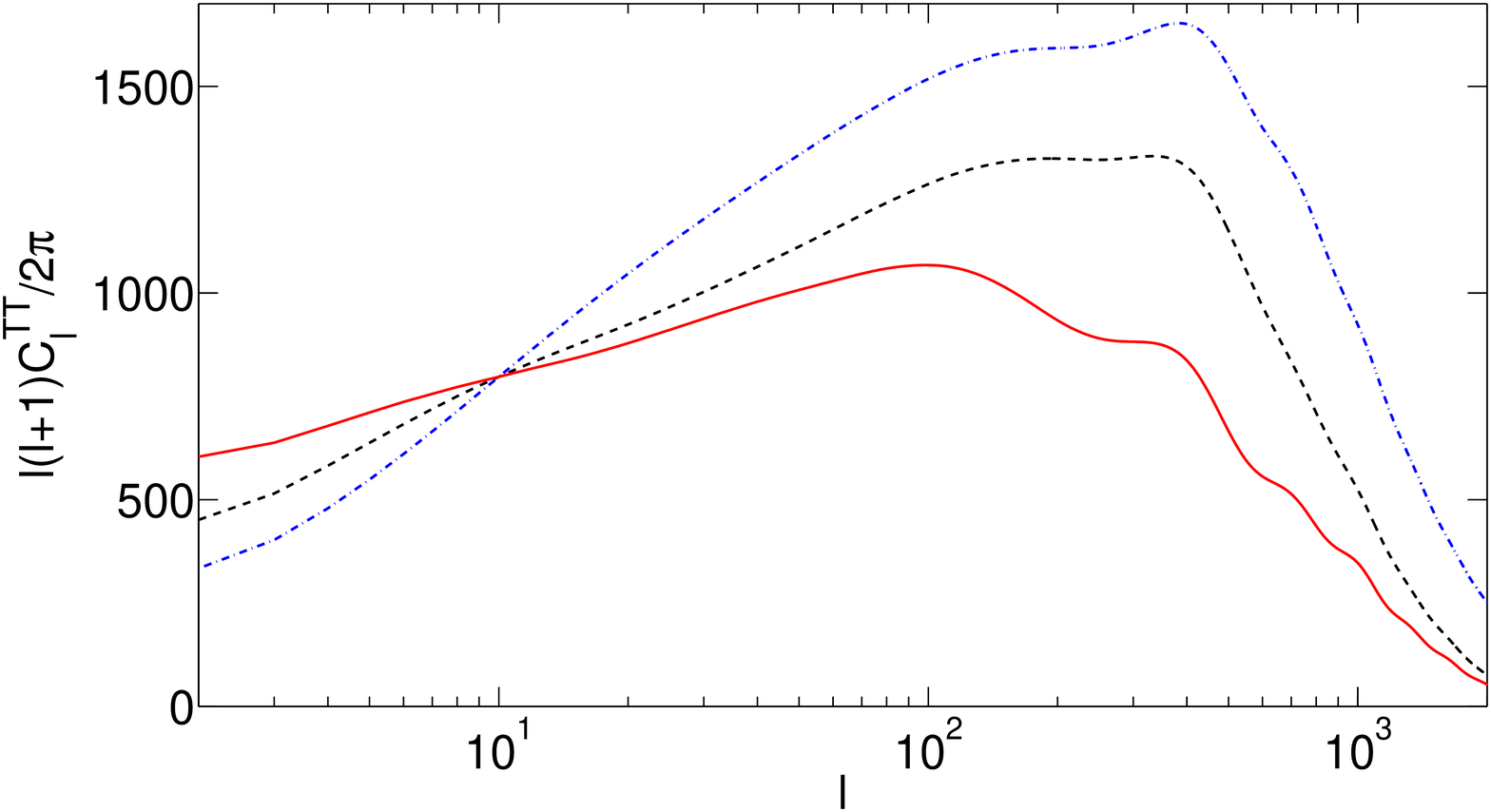}}\\
\resizebox{\columnwidth}{!}{\includegraphics{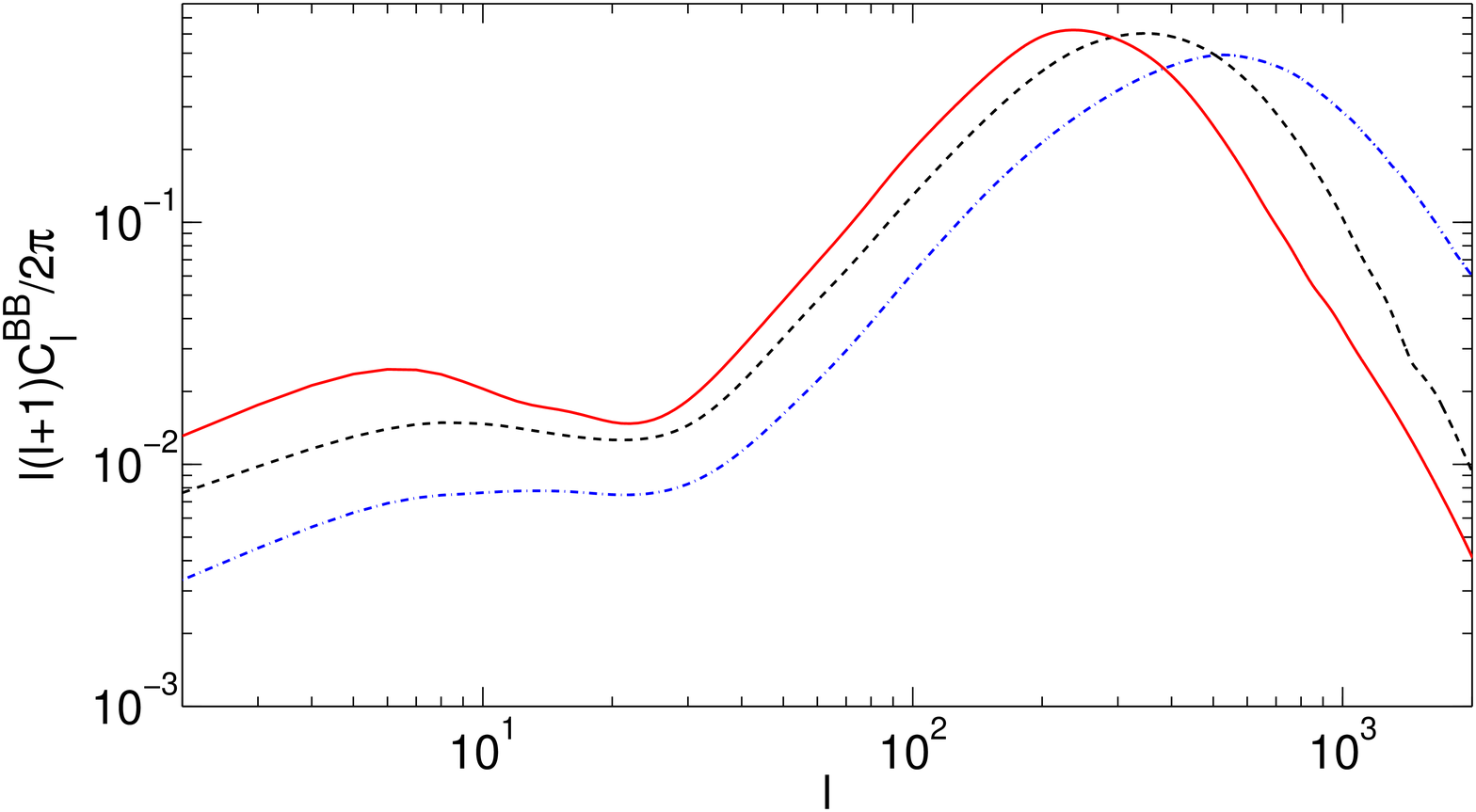}}
\caption{\label{gmu10} Temperature (TT) and B-mode polarization (BB) defect spectra, normalized to make the temperature spectra match {\it Planck} data at $\ell=10$. Different lines correspond to textures (solid red line), semilocal strings (dashed black line), and Abelian Higgs strings (dot-dashed blue line).}
\end{figure}

Figure~\ref{gmu10} shows the power spectra obtained from the field theoretical simulations of Abelian Higgs strings (AH) \cite{Bevis:2010gj}, semilocal strings (SL), and textures (TX) \cite{Urrestilla:2007sf}, for the temperature and B-mode polarization spectra, normalised to the {\it Planck} temperature power spectrum at $\ell=10$.
There are important differences between the power spectra obtained from defects or from inflation. Defects produce scalar, vector and tensor perturbations in proportions which are fixed for a given defect model, while in inflationary models vector modes are absent and the tensor contribution can vary almost independently of the scalar, apart from the inflationary consistency relation \cite{liddle:1992wi}.  In addition, defect-induced polarization is suppressed on large angular scales, as causality requires their fluctuations to be uncorrelated beyond the horizon distance at decoupling \cite{Durrer:2014raa}.

The amplitude  of the perturbations produced by defects is usually parametrised by the dimensionless number $G\mu$, where $G$ is Newton's constant and $\mu = 2\pi v^2$, where $v$ is the expectation value of the canonically-normalised symmetry-breaking field, assumed complex.\footnote{For Abelian Higgs strings at critical coupling, $\mu$ is the energy per unit length. For a theory with a canonically-normalized scalar field with expectation value $\ka$, we have $G\mu = \pi\ka^2$.
The texture literature uses a parameter $\epsilon$, defined as $\ep=4\pi G\kappa^2$ \cite{Durrer:1998rw}.  For details see the appendix in Ref.~\cite{Urrestilla:2007sf}.}  A parameter often used to quantify the contribution of defects to the power spectrum is $f_{10}$, which is the fractional contribution of
defects to the total model temperature power spectrum at multipole $\ell=10$.  With these definitions, and for small contributions from defects, $\fd \propto (G\mu)^2$.
The values of $G\mu$ needed to fit the {\em Planck} data at $\ell=10$ (i.e.\ the value for which $\fd = 1$), and the {\em Planck} 95\% upper bounds for \gmu\ and $\fd$, can be found in Table~\ref{gmut}.
Note that $G\mu_{10}$ is calculated as the normalization of strings needed to match the {\it observed} power at $\ell=10$, whereas the limit on 
$f_{10}$ is the upper bound on the ratio of the power in strings to the total power in the best-fit model at $\ell=10$.

\begin{table}
\renewcommand{\arraystretch}{1.2}
\begin{tabular}{|l|c|c|c|}
\hline
&	$G\mu_{10}$ & $G\mu$  ($<95\%$) & $\fd$ ($<95\%$)\\
\hline
Abelian Higgs strings &	$19\times 10^{-7}$ & $3.2\times 10^{-7}$ & 0.024 \\
Semilocal strings &	$53\times 10^{-7}$ & $11 \times 10^{-7}$ & 0.041\\
Textures  &	$44\times 10^{-7}$& $11 \times 10^{-7}$ & 0.054\\
\hline
\end{tabular}
\caption{\label{gmut} $G\mu_{10}$ is the normalization of different defects to match the observed $\ell=10$ multipole value (i.e.\ to explain the full temperature signal at that multipole, $f_{10} = 1$). The last two columns show the $95\%$ confidence upper limit obtained by the {\it Planck} collaboration \cite{Ade:2013xla} for the {\it Planck} + WP + {High-$\ell$} dataset.}
\end{table}

\begin{figure}[t]
\resizebox{\columnwidth}{!}{\includegraphics{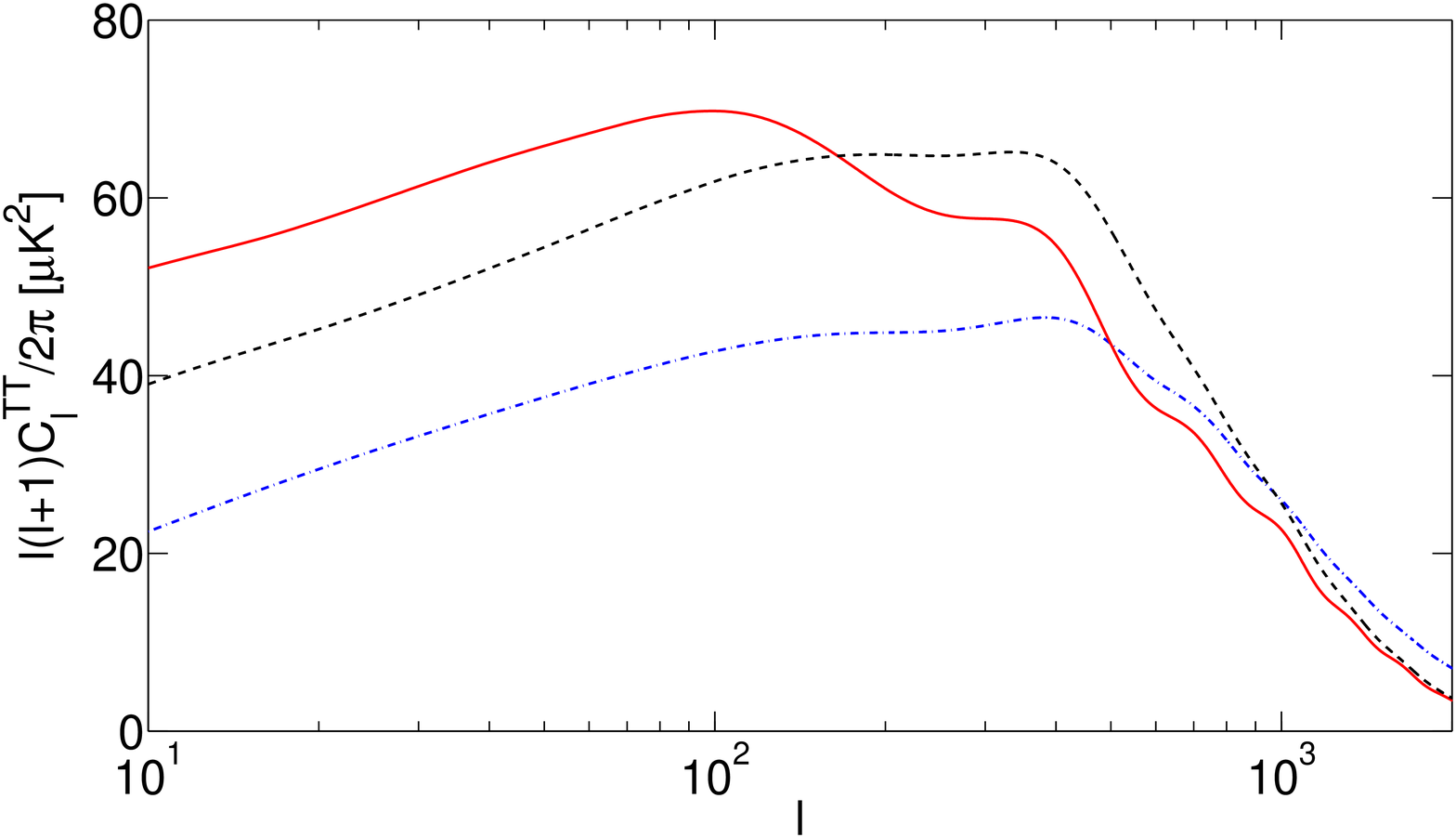}} \\
\resizebox{\columnwidth}{!}{\includegraphics{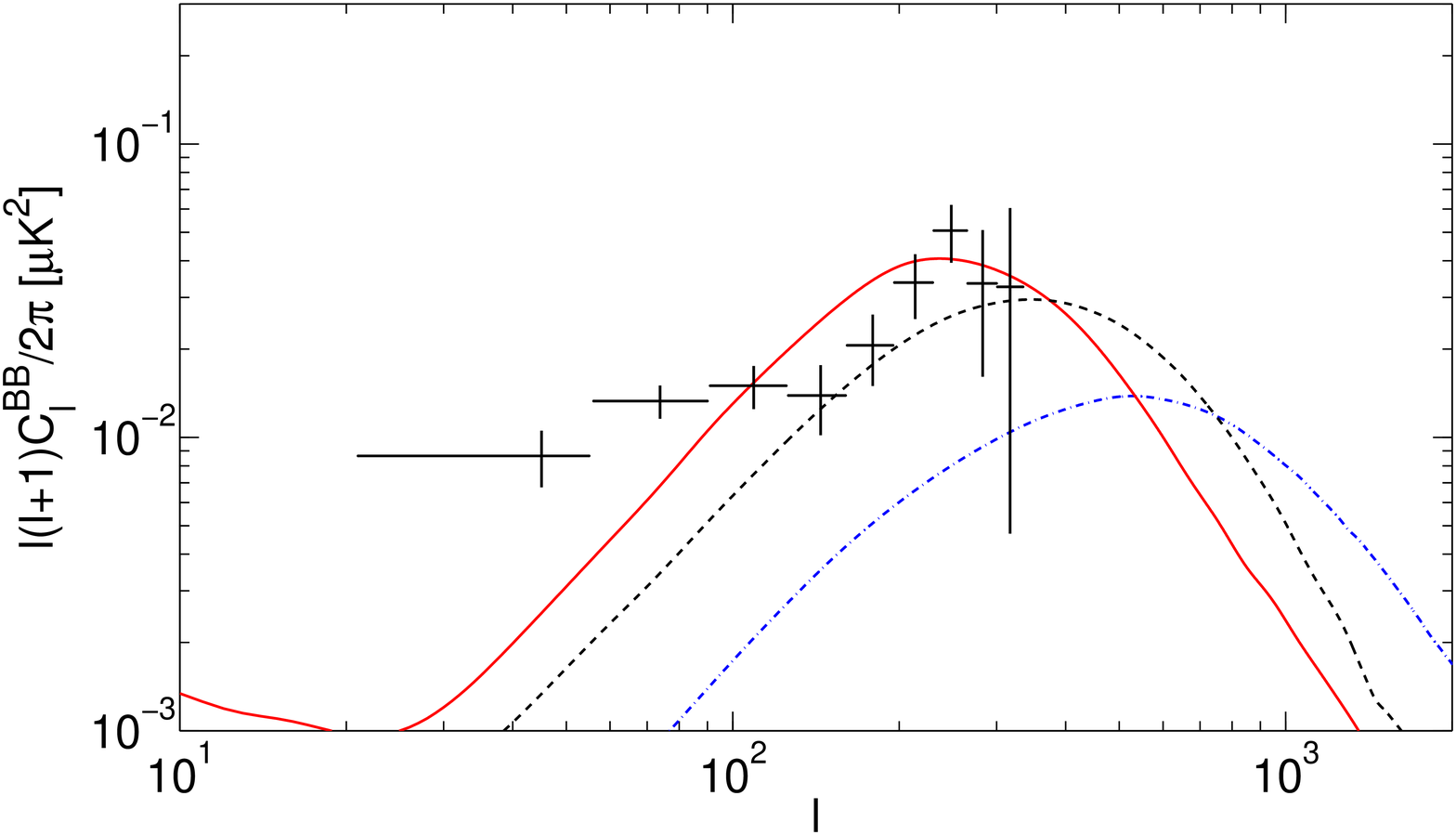}}
\caption{\label{95_Defects} Defect spectra normalised to the $95\%$ upper limits obtained using {\it Planck} + WP + High-$\ell$. Different lines correspond to textures (solid red line), semilocal strings (dashed black line), and Abelian Higgs strings (dot-dashed blue line).}
\end{figure}

CMB data already put strong constraints on the defect contribution in models which combine it with a primordial inflationary power spectrum, mainly through the increasingly accurate measurement of the temperature power spectrum to higher and higher multipoles \cite{Wyman:2005tu,Bevis:2007gh,Battye:2010xz,Dunkley:2010ge,Urrestilla:2011gr,Avgoustidis:2011ax,Ade:2013xla}. Figure~\ref{95_Defects} shows the temperature and B-mode spectra for AH, SL and TX at the upper 95\% level for defects obtained in Ref.~\cite{Ade:2013xla}. Even though the shapes of the spectra are similar (see Fig.~\ref{gmu10}) the peaks are not exactly at the same $\ell$, and they fall off at different rates at high $\ell$. 
The cosmic string model (AH) has the slowest fall-off, and its amplitude is the most tightly constrained by the temperature data.  As a result, the possible B-mode contribution is the smallest at low $\ell$ (lower panel).

As explained in Ref.~\cite{Lizarraga:2014eaa}, and as is also clear from Fig.~\ref{95_Defects}, 
the shape of the power spectrum of the defects is qualitatively wrong, and cannot give a good fit to BICEP2 data. Including  the constraint from the temperature power spectrum is likely to make the fit even worse. A similar conclusion was obtained in Ref.~\cite{Durrer:2014raa} for  self-ordering scalar fields, which is understandable since self-ordering scalar fields are closely related to the O(4) model under study here.

An apparently contradictory conclusion was obtained in Ref.\ \cite{Moss:2014cra}, where 
the BICEP2 data was fitted to the USM, allowing the inter-string distance parameter to vary. 
They found that a string-only model differed in $\chi^2$ by only 2.65 from the best-fit model with primordial tensor modes, albeit 
for inter-string distance values larger than the causal horizon at decoupling. It was suggested the model spectrum was 
representative of global strings or textures.  Numerical simulations of textures do show that they have a larger correlation length than the other defects, and hence a B-mode peak at lower $\ell$, but it is also apparent from a comparison of Fig.\ \ref{95_Defects} to Fig.\ 1 of Ref.\ \cite{Moss:2014cra}
that the shape of the texture power spectrum at low $\ell$ is not accurately modelled by the USM. 
For example, at $\ell = 70$, the best-fit USM spectrum is approximately 50\% higher than the texture spectrum, and twice as high at $\ell=40$, 
which will tend to make the texture spectrum a worse fit to the data.

We will see (Table \ref{onlyBICEP}) that, when fitting BICEP2 data only,  the $\Delta\chi^2$ between the O(4) texture model and primordial tensor modes is 5.0, significantly larger than the USM best-fit value.
We will also see that textures can combine with primordial gravitational waves to improve the fit to the BICEP2 data, 
as they help with the points at $\ell \gtrsim 200$ which are above the lensing signal \cite{Lizarraga:2014eaa,Moss:2014cra,Durrer:2014raa}.

Note that while a super-horizon inter-string distance is physically questionable, it was argued \cite{Moss:2014cra} that one could appear in models where the string-forming phase transition happens during inflation, leading to the delayed onset of scaling in the string network. However, an analysis of the delayed scaling model which takes into account the dynamics of the inter-string distance finds that a string-only model described by the USM is not a good fit \cite{Kamada:2014qta}.

\section{Models and methodology}
\label{models}

We perform a set of parameter estimations for models where  defects coexist with other sources of B-mode polarisation, namely inflationary gravitational waves, dust and lensing.
The lensing signal is always present, and was recently detected by POLARBEAR \cite{Ade:2013gez,Ade:2014afa}. However, the extra constraining power of the POLARBEAR data is weak, and for simplicity we do not include it in our analysis.

In order to reliably explore the parameter space we perform Markov Chain Monte-Carlo (MCMC) runs with the publicly-available \verb Monte \ \verb Python \ code \cite{Audren:2012wb,MontePython}, which uses \verb Class \ \cite{Lesgourgues:2011re,Blas:2011rf} as its Boltzmann equation solver for the inflationary component of the power spectrum. We compare our predictions to the
following CMB datasets:
\begin{itemize}
\item {\it Planck}+WP: {\it Planck} 2013 data \cite{Ade:2013zuv} (low-$\ell$ and high-$\ell$), including WMAP 9 year \cite{Hinshaw:2012aka} low-$\ell$ polarisation data.
\item High-$\ell$: SPT \cite{Story:2012wx, Reichardt:2011yv} and ACT \cite{Sievers:2013ica}.
\item {BICEP2}: BICEP2 BB polarization data \cite{Ade:2014xna}.
\end{itemize}
The likelihoods are the official codes provided by each experiment.\footnote{Although there may be minor differences between the likelihood
used by the BICEP2 collaboration in their publication \cite{Ade:2014xna} and the public version \cite{Audren:2014cea}, we expect that this is
not important for our main conclusions.}

The basic inflationary \LCDM\ model, the ``Power-Law'' (\PL) model, is represented by the following set of parameters:
\begin{equation}
\{ \Ochh, \Obhh, \tau, H_0, \As, \ns \}
\end{equation}
where $\Ochh$ is the physical cold dark matter density, $\Obhh$ is the baryon density, $\tau$ is the optical depth to reionization, $H_0$ is the Hubble constant, $\As$ is the amplitude of the scalar spectrum, and $\ns$ its spectral index.

We add a number  of extra ingredients  to the \PL\ model, sometimes by themselves, sometimes in combinations. Our main extra ingredient is given by topological defects parametrized as $(G\mu)^2$, for each of the three models explained in the previous section (Abelian Higgs cosmic strings AH, textures TX, or semilocal strings SL). Another parameter describing an extra ingredient is $r$ which parametrises the amount of inflationary gravitational waves through the tensor-to-scalar ratio (at $k=0.002 \, \mathrm{Mpc^{-1}}$). Scalar perturbation quantities are also specified at a pivot scale $k=0.002 \, \mathrm{Mpc^{-1}}$.

\begin{table*}[t]
\begin{center}
\renewcommand{\arraystretch}{1.2}
\begin{tabular}{|c||c|c|c||c|c|c||c|} 
\hline
Dataset &  \multicolumn{7}{c|}{BICEP2 (only {\it BB})} \\
\hline
Model & \multicolumn{3}{c||}{\PLgmu} &  \multicolumn{3}{c||}{\PLrgmu} & \PLr \\ \hline
 Param & AH  & SL & TX & AH  & SL & TX& -\\ \hline
$r$ & - & - & - & $0.14_{-0.06}^{+0.04}$ &  $0.14_{-0.06}^{+0.04}$& $0.14_{-0.06}^{+0.04}$   &$0.21_{-0.05}^{+0.04}$\\ 
$10^{12}(G\mu)^2$ &   $0.40_{-0.08}^{+0.07}$& $1.73_{-0.32}^{+0.29}$   & $0.86_{-0.16}^{+0.14}$ & $0.20_{-0.09}^{+0.08}$& $0.87_{-0.39}^{+0.34}$& $0.43_{-0.20}^{+0.17}$  & -\\ 
\hline 
$-\ln{\cal L}_\mathrm{max}$ & $8.1$ & $7.4$ & $6.8$ & $1.6$ & $1.6$ & $1.8$  & $4.3$ \\\hline
 \end{tabular} \\ 
 \caption{\label{onlyBICEP} Parameter estimations and best-fit likelihood values for various cosmological models, fitting for the BICEP2 data. Only the B-mode is used for these estimations.}
\end{center} 
\end{table*}  

\begin{table*}[t]
\begin{center}
\renewcommand{\arraystretch}{1.2}
\begin{tabular}{|c||c|c|c||c|c|c||c|} 
\hline
Dataset &  \multicolumn{7}{c|}{{\it Planck} + WP + High-$\ell$ + BICEP2} \\
\hline
Model & \multicolumn{3}{c||}{\PLgmu} & \multicolumn{3}{c||}{\PLrgmu} & \PLr \\ \hline
 Param & AH  & SL & TX & AH  & SL & TX& -\\ \hline
$n_{s }$ & $0.955_{-0.008}^{+0.007}$  & $0.964_{-0.008}^{+0.007}$& $0.962_{-0.007}^{+0.007}$ & $0.963_{-0.008}^{+0.007}$& $0.966_{-0.008}^{+0.008}$ & $0.965_{-0.007}^{+0.007}$  & $0.962_{-0.007}^{+0.007}$ \\ 
$r$ & - & - & - & $0.14_{-0.04}^{+0.03}$ & $0.10_{-0.04}^{+0.03}$ & $0.09_{-0.04}^{+0.03}$  & $0.15_{-0.04}^{+0.03}$\\ 
$10^{12}(G\mu)^2$ &  $0.084_{-0.025}^{+0.026}$  & $1.34_{-0.29}^{+0.27}$ & $0.73_{-0.15}^{+0.14}$ &  $ < 0.083$ & $ < 1.3$ & $ < 0.74$  &  -\\ 
$f_{10 }$ &   $0.021_{-0.006}^{+0.006}$   &  $0.044_{-0.010}^{+0.009}$ &  $0.035_{-0.007}^{+0.006}$ &  $ < 0.020$  & $ < 0.042$ &  $ < 0.035$  & - \\ 
\hline 
$-\ln{\cal L}_\mathrm{max}$ & $5280.1$  & $5268.8$ & $5266.8$ & $5264.4$ & $5262.9$ & $5263.0$  & $5265.8$ \\\hline
 \end{tabular} \\ 
 \caption{\label{allCMB} Parameter estimations and best-fit likelihood values for various cosmological models, fitting for all the CMB data.}
\end{center} 
\end{table*}

The BICEP2 collaboration included also the running of the scalar spectral index $\alpha_{\rm s}$ in order to improve the agreement between the BICEP2 and {\it Planck} data \cite{Ade:2014xna}. Although several papers \cite{Audren:2014cea,Smith:2014kka,Martin:2014lra}, showed that there is no worrying tension between BICEP2 and {\it Planck} data, we nevertheless also study the impact of $\alpha_{\rm s}$ here.

As mentioned above, the observed B-mode polarization signal may have a contribution from polarised dust emission  \cite{Mortonson:2014bja,Flauger:2014qra}. We characterized this B-mode channel  by $A_{\rm dust}$ \footnote{Another parametrisation of dust is used in the literature, given by $\Delta^2_{BB}$, which is related to ours via $$\Delta^2_{BB,{\rm dust},\ell}=\frac{\ell^2}{2\pi} C_\ell=\frac{A_{\rm dust}}{2\pi}\ell^{-0.3}$$}, using the dust  model proposed by the {\it Planck} collaboration \cite{PlanckDust}:
\begin{equation}
C_\ell^{BB,{\rm dust}} = A_{\rm dust} \ell^{-2.3}
\label{dustlaw}
\end{equation}
Our models are constructed using those building blocks, starting from the models with just one extra ingredient, and moving to more complex models where several additional ingredients are present simultaneously.

\section{Results}
\label{results}

In this section we present the results from fitting different combinations of datasets with various cosmological models. 

First, in subsection~\ref{rgmu} we fit CMB data with our basic model (\PL)
with defects $G\mu$, with inflationary tensor modes $r$, and with both $r$ and $G\mu$. The CMB data chosen are the BICEP2 data alone (for which we only fit for the B-mode spectrum), or all the CMB data.  We also considered the case where the data used did not include the High-$\ell$ data, but 
the results for both these two choices of data (with and without High-$\ell$) were identical, so we only show the parameter constraints with all the CMB data.
In subsection~\ref{alphas} we consider a model where the running of the scalar spectral index $\alpha_{\rm s}$ is also free. The last case, subsection~\ref{dust}, corresponds to models which include a dust contribution as described above.

The results showed in the Tables in the subsequent sections state only the relevant parameters for the given case. In all cases flat parameter priors were used, in the case of defects the prior being flat in $(G\mu)^2$ which is proportional to the fractional defect contribution to the power spectra $\fd$. 
The prior ranges were
\mbox{$0 < 10^{12} (G\mu)^2 < 4$} and \mbox{$0  < A_{\rm dust}/(\mu K)^2 <0.75$}.
All other parameters, including foreground parameters with the exception of the new polarised dust amplitude, were modelled as in the {\it Planck} collaboration papers \cite{Ade:2013kta,Ade:2013zuv}.

\subsection{Primordial tensor modes and defects}
\label{rgmu}

We begin the analysis by extending the results of our previous paper \cite{Lizarraga:2014eaa} with quantitative statements. All results, using only BICEP2 BB data and using the full CMB set, can be found in Tables \ref{onlyBICEP} and \ref{allCMB} respectively. The structure of the tables is the following: on the left we show results from  chains containing defects; on the right-hand side, we show  results from models without defects, included here as reference values.

The values in  Table~\ref{onlyBICEP}, especially best-fit likelihoods,  
show that the fit is rather poor, as  suggested in Ref.~\cite{Lizarraga:2014eaa}. 
Actually, for a model with only one extra component,  none of 
the defect models (\PLgmu) provides a fit that is comparable to the model including only inflationary gravitational waves (\PLr), although the texture fit is only moderately worse.

We then fit the BICEP2 data with a model which contains both defects and gravitational waves (\PLrgmu), in order to assess whether defects could assist tensor modes. As mentioned before, at low $\ell$ defects cannot explain the power measured. Nevertheless, defects peak at higher $\ell$, which might help to fit those points that lie above the lensing curve. The fit is improved  (the likelihood is better), although it should be noted that  this last model has 2 extra ingredients.

As a next step we use the full CMB dataset ({\it Planck} + WP + High-$\ell$ + BICEP2) and include the contributions to temperature and polarization (both E- and B-modes)  from the different ingredients. If we compare models with only one extra ingredient, we find that \PLAH\ fits the data quite poorly, whereas \PLr, \PLTX\ and \PLSL\ fit the data at roughly the same level, with $r$ being the best model followed closely by TX.

The \gmu\ constraint obtained from the full set of CMB data is tighter than that from only BICEP2, especially for the AH case. For this case, {\it Planck} bounds are strong enough to push the corresponding BB spectrum far below the BICEP2 data, in other words, the BICEP2 data do not constrain further the AH model in the combined {\it Planck} + BICEP2 case.
By contrast, temperature bounds for SL and TX  \cite{Ade:2013xla} leave their BB power spectra around the values of BICEP2 (for high $\ell$), such that BICEP2 alone is able to put comparable constraints on the level of allowed defects. Our results are consistent with the observation that accurately-determined B-modes can distinguish between different types of defects \cite{Mukherjee:2010ve}.

The final possibility is the mixture of inflationary gravitational waves and cosmic defects, \PLrgmu.  We observe that there is less room for defects, and we only obtain upper 95\% bounds. Roughly, the mean values in the previous cases become 95\% values now. Here again SL and TX do marginally better than AH strings, which are not able to lower $r$. The reason has already been mentioned: their contribution is so suppressed by the constraints from the {\it Planck} data that the effect on tensor modes is negligible.

\subsection{Running of the scalar index}
\label{alphas}

Here we do not consider the BICEP2 data alone, since the running of the scalar spectral index affects mainly the temperature channel. The results from a fit to the full CMB dataset can be found in Table~\ref{running}.

The model which contains gravitational waves plus running is slightly preferred over \PLrgmu, possibly because if one also allows for running, $r$ could take higher values (e.g.\ see \PLralpha\ of Table \ref{running}) and therefore a better fit of B-modes. Another typical effect of including $\alpha_{\rm s}$ is that the scalar spectral index is pushed up, which in principle implies more room for a defect contribution, though in this case it only affects AH strings. Running also changes the tilt of the temperature spectrum, causing an unexpected anticorrelation between \gmu\ and $\fd$.

It is worth noting that for the TX and SL cases, allowing for the running of the scalar index does not increase the value of $r$; it remains around the same values as for cases without $\alpha_{\rm s}$. At the same time, allowing for defects does not reduce the magnitude of the running.

\begin{table*}
\begin{center}
\renewcommand{\arraystretch}{1.2}
\begin{tabular}{|c||c|c|c||c|} 
\hline
Dataset &  \multicolumn{4}{c|}{{\it Planck} + WP + High-$\ell$ + BICEP2} \\
\hline
Model & \multicolumn{3}{c||}{\PLrgmualpha } & \PLralpha \\ \hline
 Param & AH  & SL& TX & - \\ \hline
 $n_{\rm s}$ & $1.061_{-0.028}^{+0.028}$ & $1.048_{-0.031}^{+0.026}$ & $1.049_{-0.032}^{+0.030}$& $1.055_{-0.031}^{+0.031}$ \\
 $\alpha_{\rm s}$ & $-0.032_{-0.009}^{+0.008}$  & $-0.027_{-0.008}^{+0.010}$ & $-0.027_{-0.010}^{+0.010}$ & $-0.030_{-0.010}^{+0.010}$ \\
 $r$ & $0.19_{-0.05}^{+0.04}$ & $0.15_{-0.06}^{+0.04}$ & $0.14_{-0.06}^{+0.04}$    & $0.20_{-0.05}^{+0.04}$\\
$10^{12}(G\mu)^2$ &  $ < 0.10$ & $ < 1.3$ & $ < 0.71$& -  \\ 
$f_{10 }$ &   $ < 0.030$  & $ < 0.047$ & $ < 0.038$ & - \\ 
\hline 
$-\ln{\cal L}_\mathrm{max}$ & $5260.3$& $5259.2$ & $5258.6$ & $5261.6$ \\\hline
 \end{tabular} \\ 
 \caption{\label{running} Parameter estimations and best-fit likelihood values for different cosmological models, fitting for all the CMB data. These cosmological modes allow for the running of the scalar index.}
\end{center}
\end{table*}

\subsection{Dust}
\label{dust}

As discussed in Section~\ref{models},  we consider a dust model proposed by the {\it Planck} collaboration \cite{PlanckDust}, given by 
\begin{equation}
C_\ell^{BB,{\rm dust}} = A_{\rm dust} \ell^{-2.3}
\end{equation}
A similar model has been used  by Mortonson and Seljak~\cite{Mortonson:2014bja}  and Flauger {\it et al.} \cite{Flauger:2014qra} to examine the robustness of the BICEP2 result's interpretation as primordial, and we follow their approach.

\begin{figure}[b]
\resizebox{\columnwidth}{!}{\includegraphics{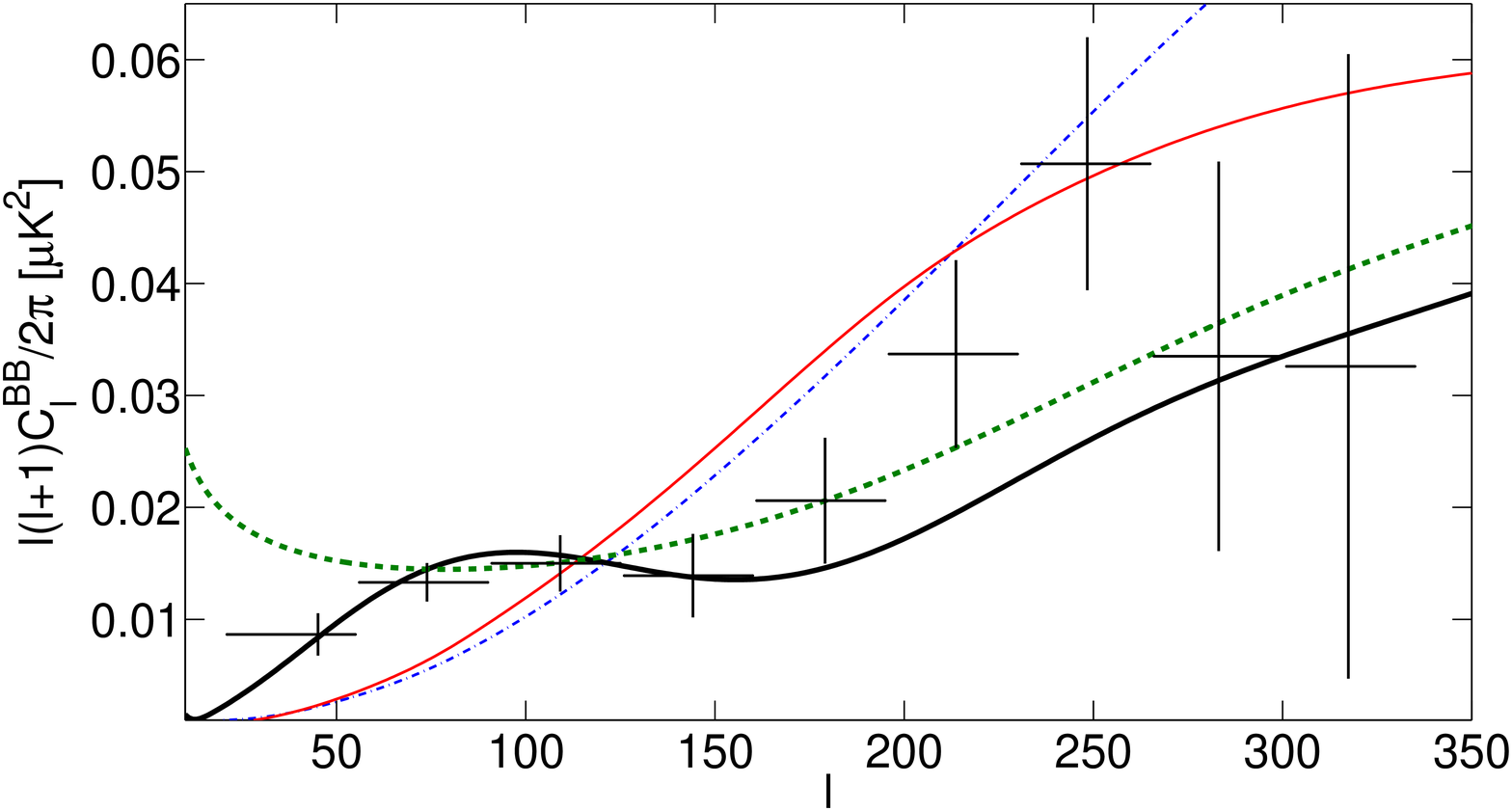}} \\
\caption{\label{rGmuDust_Normalized} B-mode spectra, including the lensing contribution, using best-fit normalization values given in Tables~\ref{onlyBICEP} and \ref{onlyBICEPDust}, for tensors (thick solid black line), dust (thick dashed green line), AH strings (dot-dashed blue) and textures (solid red line). BICEP2 data points are also shown.}
\end{figure}

In Fig.~\ref{rGmuDust_Normalized} we show the contributions to the B-mode power spectrum from inflationary tensors, AH strings, textures, and dust, together with the data points from BICEP2. The normalization is the one obtained from fitting only the BICEP2 data to a model \PL\ plus one extra ingredient (see Tables~\ref{onlyBICEP} and \ref{onlyBICEPDust}). Note that the lensing spectrum is added in all  cases. In the figure it can be seen that dust and $r$ have more importance for lower $\ell$; therefore, in B-modes dust is in more direct competition with $r$ than with defects.

As  in the previous section, we start analyzing BICEP2 B-mode data. The first thing we notice is that a model including just dust as an extra ingredient is able to improve the fit of \PLr. Moreover, if we consider a composite model (\PLrdust), the gravitational wave detection disappears and  the fit does not improve (as previously found in Ref.~\cite{Mortonson:2014bja}). The best-fit value for $r$ is at $r=0$, i.e.\ a model with dust alone provides the best fit.

Dust combined with defects gives better results than with gravitational waves. Note that in all cases (be it with $r$ or with any $G\mu$) the dust contribution is at the same level. However, dust lowers the amount of defects to about half the one obtained using \PLgmu, and more or less at the level of \PLrgmu. This last set of models does not improve the best-fit likelihood. Finally, in a model with all ingredients (\PLrgmudust), we find that a model with no dust is possible at one-sigma, and thus we quote an upper $95\%$ confidence limit. This is due to the fact that dust and inflationary tensors can both account for the low $\ell$ part of the spectrum, whereas defects account for the higher $\ell$. 

Considering the full CMB dataset, the picture is roughly the same. Dust does a very good job on its own, and any other combination improves only marginally the best-fit log-likelihood.  Once again, since the temperature power spectrum is also constraining the defect contribution, we  find only 95\% upper bounds for defects. The bounds for SL and especially TX are tighter than those from {\it Planck} (see Table~\ref{gmut}).  In all cases, the combination \PLgmudust\ does better than the equivalent \PLrgmu.

Note that the different mean values of $A_{\rm dust}$ are due to the differences in the lensing spectra due to different cosmologies used in Tables~\ref{onlyBICEPDust} and~\ref{allCMBDust}.

\begin{table*}
\begin{center}
\renewcommand{\arraystretch}{1.2}
\begin{tabular}{|c||c|c|c||c|c|c||c||c|} 
\hline
Dataset &  \multicolumn{8}{c|}{BICEP2 (only {\it BB})} \\
\hline
Model & \multicolumn{3}{c||}{\PLgmudust} &  \multicolumn{3}{c||}{\PLrgmudust} & \PLrdust &  \PLdust \\ \hline
 Param & AH & SL & TX & AH & SL & TX  & - & -\\ \hline
$r$ & - & - & - &  $< 0.11$  &  $< 0.10$ &  $< 0.18$ &$< 0.22$ & - \\ 
$10^{12}(G\mu)^2$ &   $0.17_{-0.10}^{+0.08}$  & $0.74_{-0.40}^{+0.40}$ & $0.37_{-0.24}^{+0.16}$ & $0.17_{-0.08}^{+0.05}$ & $0.72_{-0.41}^{+0.32}$ & $0.36_{-0.21}^{+0.16}$  & - & -\\ 
$A_{\rm dust} \,[\mu K^2]$ & $0.20_{-0.08}^{+0.06}$ & $0.20_{-0.08}^{+0.06}$ & $0.19_{-0.09}^{+0.06}$  & $< 0.26$  & $< 0.25$  & $< 0.25$ & $0.19_{-0.10}^{+0.10}$  & $0.30_{-0.07}^{+0.06}$ \\
\hline 
$-\ln{\cal L}_\mathrm{max}$ & $1.7$ & $1.7$  & $1.8$  & $1.5$& $1.5$ & $1.7$  & $3.3$ & $3.3$ \\\hline
 \end{tabular} \\ 
  \caption{\label{onlyBICEPDust} Parameter estimations and best-fit likelihood values for different cosmological models, fitting for the BICEP2 data. This is similar to Table~\ref{onlyBICEP}, but in this case a dust model is included.}
\end{center} 
\end{table*}

\begin{table*}
\begin{center}
\renewcommand{\arraystretch}{1.2}
\begin{tabular}{|c||c|c|c||c|c|c||c||c|} 
\hline
Dataset &  \multicolumn{8}{c|}{{\it Planck} + WP + High-$\ell$ + BICEP2} \\
\hline
Model & \multicolumn{3}{c||}{\PLgmudust} &  \multicolumn{3}{c||}{\PLrgmudust} & \PLrdust &  \PLdust \\ \hline
 Param & AH  & SL & TX& AH  & SL & TX& - & -\\ \hline
$r$  &- & -  & - & $< 0.10$  & $< 0.09$ & $< 0.10$ &  $< 0.11$ & - \\ 
$10^{12}(G\mu)^2$ & $< 0.074$  & $< 0.97$ & $< 0.54$ &  $< 0.075$  & $< 0.97$ & $< 0.56$ & - & - \\ 
$f_{10 }$ & $< 0.019$  & $< 0.031$ & $< 0.026$ &  $< 0.019$  & $ < 0.032$  &  $< 0.027$ & - & - \\
\rule{0pt}{12pt}$A_{\rm dust} \,[\mu K^2]$& $0.24_{-0.06}^{+0.05}$ & $0.21_{-0.06}^{+0.06}$ & $0.19_{-0.07}^{+0.06}$ & $0.20_{-0.07}^{+0.06}$ & $0.16_{-0.07}^{+0.07}$ & $0.15_{-0.07}^{+0.06}$  &  $0.21_{-0.06}^{+0.07}$ & $0.25_{-0.06}^{+0.05}$\\ 
\hline 
$-\ln{\cal L}_\mathrm{max}$ & $5259.6$ & $5259.4$  & $5258.5$ & $5259.5$ & $5259.3$ & $5258.4$ & $5259.6$ & $5260.0$ \\\hline
 \end{tabular} \\ 
  \caption{\label{allCMBDust} Parameter estimations and best-fit likelihood values for different cosmological models, fitting for all the CMB data. This is similar to Table~\ref{allCMB}, but in this case a dust model is included.}\end{center} 
\end{table*}

\section{Discussion and conclusions}

In this paper we investigate quantitatively the impact of the recent detection of B-mode polarisation by the BICEP2 collaboration \cite{Ade:2014xna}
on models containing topological defects, extending the results of our earlier, more qualitative, study \cite{Lizarraga:2014eaa}. In accordance with
our earlier paper, we find that topological defects on their own are a poor fit to the signal and that we need an additional contribution, either inflationary
gravitational waves or polarised dust emission.

When considering only the BICEP2 B-mode polarization data, we find that the combination of topological defects and inflationary gravitational waves
(\PLrgmu) or topological defects and dust (\PLgmudust) slightly improves the fit over inflationary gravitational waves alone (\PLr) or dust alone (\PLdust). This is because topological defects help to fit the BICEP2 data points
at $\ell \gtrsim 200$, which lie above the lensing B-mode contribution, cf.\ Fig.\ \ref{rGmuDust_Normalized}. The combination of inflationary gravitational waves
and dust on the other hand does not improve the fit over either contribution alone.
We note that there are hints in the cross-correlation between BICEP2 and Keck array data  \cite{Ade:2014xna} that the central values of the B-mode power spectrum will decrease in the future, which will have the effect of more strongly constraining the defect contribution.

The situation changes slightly when we consider the full CMB dataset, consisting of {\it Planck} + WP + High-$\ell$ + BICEP2. In this case the texture
model on its own (\PLTX) is only slightly worse than inflationary gravitational waves (\PLr), while cosmic strings 
(\PLAH) are ruled out as the sole source of B-modes. Dust on the other hand is much better, so that \PLdust\ is the globally-preferred model, and neither defects nor inflationary gravitational waves are able to improve the goodness of fit significantly.

When considering parameter constraints on $G\mu$, we find that BICEP2 on its own constrains the SL and TX models to roughly the same level as {\it Planck} data do. In other words, the constraints obtained from BICEP2 alone for SL and TX are as strong as the ones obtained from {\it Planck} data. On the other hand, {\it Planck} data constrains AH strings more strongly than BICEP2. The reason is that the combined temperature anisotropy dataset constraints on defects come from $\ell \gtrsim 100$, where the string-induced power spectrum peaks higher and decays more slowly than those of the other defects for a given $G\mu$. 

\begin{table}
\renewcommand{\arraystretch}{1.2}
\begin{tabular}{|l|c|c|}
\hline
&	 $G\mu$ at $<95\%$ & $\fd$ at $<95\%$\\
\hline
Abelian Higgs strings &$2.7\times 10^{-7}$  & 0.019\\
Semilocal strings & $9.8 \times 10^{-7}$& 0.031\\
Textures  & $7.3 \times 10^{-7}$  & 0.026\\
\hline
\end{tabular}
\caption{\label{gmu} $95\%$ confidence limits for \gmu\ and $\fd$ obtained using \PLgmudust\ and the full CMB dataset.}
\end{table}

The constraints become tighter when including a contribution from inflationary gravitational waves or dust.
For the full CMB dataset and for a model with defects and dust
 (see Table \ref{gmu}), we find $G\mu < 2.7 \times 10^{-7}$ for AH, $G\mu < 7.3 \times 10^{-7}$ for TX and $G\mu < 9.8 \times 10^{-7}$ for SL (all at 95\%).
These constraints are tighter
than ones found by the {\it Planck} collaboration for the temperature data alone \cite{Ade:2013xla} (especially for texture, see Table~\ref{gmut}), which shows the importance that even the current B-mode polarisation data has for constraining topological defects.

{\it Note added - } While this paper was being refereed the {\it Planck} collaboration submitted a paper \cite{Adam:2014bub} where they update their dust model to $C^{BB}_\ell\propto\ell^{-2.45}$ (to be compared with equation (\ref{dustlaw})). We do not expect our results to change significantly with this new power law. We tested the case of  \PLAHgmudust\ fitted to the full CMB dataset, and found that the upper 95\% confidence limit in $G\mu$  moves from $2.7\times 10^{-7}$ to $2.9\times 10^{-7}$ (see Table~\ref{gmu}), which supports our expectations.

\begin{acknowledgments}
This work has been possible thanks to the computing infrastructure of the i2Basque academic network, the COSMOS Consortium supercomputer (within the DiRAC Facility jointly funded by STFC and the Large Facilities Capital Fund of BIS), and the Andromeda cluster of the University of Geneva.
JL and JU acknowledge support from
the Basque Government (IT-559-10), the University of the Basque Country UPV/EHU (EHUA 12/11), MINECO  (FPA2012-34456) and Consolider Ingenio  (CPAN CSD2007-00042 and  EPI CSD2010-00064). DD and MK acknowledge financial support from the Swiss National Science Foundation.
MH and ARL acknowledge support 
from the Science and Technology Facilities Council (grant
numbers ST/J000477/1, ST/K006606/1, and ST/L000644/1).
\end{acknowledgments}

\vspace{3mm}

\bibliography{CosmicStrings.bib}

\begin{thebibliography}{88}
\expandafter\ifx\csname natexlab\endcsname\relax\def\natexlab#1{#1}\fi
\expandafter\ifx\csname bibnamefont\endcsname\relax
  \def\bibnamefont#1{#1}\fi
\expandafter\ifx\csname bibfnamefont\endcsname\relax
  \def\bibfnamefont#1{#1}\fi
\expandafter\ifx\csname citenamefont\endcsname\relax
  \def\citenamefont#1{#1}\fi
\expandafter\ifx\csname url\endcsname\relax
  \def\url#1{\texttt{#1}}\fi
\expandafter\ifx\csname urlprefix\endcsname\relax\def\urlprefix{URL }\fi
\providecommand{\bibinfo}[2]{#2}
\providecommand{\eprint}[2][]{\url{#2}}

\bibitem[{\citenamefont{Ade et~al.}(2014{\natexlab{a}})}]{Ade:2013gez}
\bibinfo{author}{\bibfnamefont{P.}~\bibnamefont{Ade}} \bibnamefont{et~al.}
  (\bibinfo{collaboration}{POLARBEAR Collaboration}),
  \bibinfo{journal}{Phys.Rev.Lett.} \textbf{\bibinfo{volume}{113}},
  \bibinfo{pages}{021301} (\bibinfo{year}{2014}{\natexlab{a}}),
  \eprint{1312.6646}.

\bibitem[{\citenamefont{Ade et~al.}(2014{\natexlab{b}})}]{Ade:2014afa}
\bibinfo{author}{\bibfnamefont{P.}~\bibnamefont{Ade}} \bibnamefont{et~al.}
  (\bibinfo{collaboration}{POLARBEAR Collaboration}),
  \bibinfo{journal}{Astrophys.J.} \textbf{\bibinfo{volume}{794}},
  \bibinfo{pages}{171} (\bibinfo{year}{2014}{\natexlab{b}}),
  \eprint{1403.2369}.

\bibitem[{\citenamefont{Ade et~al.}(2014{\natexlab{c}})}]{Ade:2014xna}
\bibinfo{author}{\bibfnamefont{P.}~\bibnamefont{Ade}} \bibnamefont{et~al.}
  (\bibinfo{collaboration}{BICEP2 Collaboration}), \bibinfo{journal}{{Phys.
  Rev. Lett.}} \textbf{\bibinfo{volume}{{112}}}, \bibinfo{pages}{{241101}}
  (\bibinfo{year}{2014}{\natexlab{c}}), \eprint{1403.3985}.

\bibitem[{\citenamefont{Lizarraga et~al.}(2014)\citenamefont{Lizarraga,
  Urrestilla, Daverio, Hindmarsh, Kunz, and Liddle}}]{Lizarraga:2014eaa}
\bibinfo{author}{\bibfnamefont{J.}~\bibnamefont{Lizarraga}},
  \bibinfo{author}{\bibfnamefont{J.}~\bibnamefont{Urrestilla}},
  \bibinfo{author}{\bibfnamefont{D.}~\bibnamefont{Daverio}},
  \bibinfo{author}{\bibfnamefont{M.}~\bibnamefont{Hindmarsh}},
  \bibinfo{author}{\bibfnamefont{M.}~\bibnamefont{Kunz}}, \bibnamefont{and}
  \bibinfo{author}{\bibfnamefont{A.~R.} \bibnamefont{Liddle}},
  \bibinfo{journal}{Phys.Rev.Lett.} \textbf{\bibinfo{volume}{112}},
  \bibinfo{pages}{171301} (\bibinfo{year}{2014}), \eprint{1403.4924}.

\bibitem[{\citenamefont{Moss and Pogosian}(2014)}]{Moss:2014cra}
\bibinfo{author}{\bibfnamefont{A.}~\bibnamefont{Moss}} \bibnamefont{and}
  \bibinfo{author}{\bibfnamefont{L.}~\bibnamefont{Pogosian}},
  \bibinfo{journal}{Phys.Rev.Lett.} \textbf{\bibinfo{volume}{112}},
  \bibinfo{pages}{171302} (\bibinfo{year}{2014}), \eprint{1403.6105}.

\bibitem[{\citenamefont{Durrer et~al.}(2014)\citenamefont{Durrer, Figueroa, and
  Kunz}}]{Durrer:2014raa}
\bibinfo{author}{\bibfnamefont{R.}~\bibnamefont{Durrer}},
  \bibinfo{author}{\bibfnamefont{D.~G.} \bibnamefont{Figueroa}},
  \bibnamefont{and} \bibinfo{author}{\bibfnamefont{M.}~\bibnamefont{Kunz}},
  \bibinfo{journal}{JCAP} \textbf{\bibinfo{volume}{1408}}, \bibinfo{pages}{029}
  (\bibinfo{year}{2014}), \eprint{1404.3855}.

\bibitem[{\citenamefont{Bonvin et~al.}(2014)\citenamefont{Bonvin, Durrer, and
  Maartens}}]{Bonvin:2014xia}
\bibinfo{author}{\bibfnamefont{C.}~\bibnamefont{Bonvin}},
  \bibinfo{author}{\bibfnamefont{R.}~\bibnamefont{Durrer}}, \bibnamefont{and}
  \bibinfo{author}{\bibfnamefont{R.}~\bibnamefont{Maartens}},
  \bibinfo{journal}{Phys.Rev.Lett.} \textbf{\bibinfo{volume}{112}},
  \bibinfo{pages}{191303} (\bibinfo{year}{2014}), \eprint{1403.6768}.

\bibitem[{\citenamefont{Mortonson and Seljak}(2014)}]{Mortonson:2014bja}
\bibinfo{author}{\bibfnamefont{M.~J.} \bibnamefont{Mortonson}}
  \bibnamefont{and} \bibinfo{author}{\bibfnamefont{U.}~\bibnamefont{Seljak}}
  (\bibinfo{year}{2014}), \eprint{1405.5857}.

\bibitem[{\citenamefont{Flauger et~al.}(2014)\citenamefont{Flauger, Hill, and
  Spergel}}]{Flauger:2014qra}
\bibinfo{author}{\bibfnamefont{R.}~\bibnamefont{Flauger}},
  \bibinfo{author}{\bibfnamefont{J.~C.} \bibnamefont{Hill}}, \bibnamefont{and}
  \bibinfo{author}{\bibfnamefont{D.~N.} \bibnamefont{Spergel}},
  \bibinfo{journal}{JCAP} \textbf{\bibinfo{volume}{1408}}, \bibinfo{pages}{039}
  (\bibinfo{year}{2014}), \eprint{1405.7351}.

\bibitem[{\citenamefont{Vilenkin and Shellard}(1994)}]{VilShe94}
\bibinfo{author}{\bibfnamefont{A.}~\bibnamefont{Vilenkin}} \bibnamefont{and}
  \bibinfo{author}{\bibfnamefont{E.}~\bibnamefont{Shellard}},
  \emph{\bibinfo{title}{{Cosmic strings and other topological defects}}}
  (\bibinfo{publisher}{{Cambridge University Press}}, \bibinfo{year}{1994}).

\bibitem[{\citenamefont{Hindmarsh and Kibble}(1995)}]{Hindmarsh:1994re}
\bibinfo{author}{\bibfnamefont{M.}~\bibnamefont{Hindmarsh}} \bibnamefont{and}
  \bibinfo{author}{\bibfnamefont{T.}~\bibnamefont{Kibble}},
  \bibinfo{journal}{Rept.Prog.Phys.} \textbf{\bibinfo{volume}{58}},
  \bibinfo{pages}{477} (\bibinfo{year}{1995}), \eprint{hep-ph/9411342}.

\bibitem[{\citenamefont{Durrer et~al.}(2002)\citenamefont{Durrer, Kunz, and
  Melchiorri}}]{Durrer:2001cg}
\bibinfo{author}{\bibfnamefont{R.}~\bibnamefont{Durrer}},
  \bibinfo{author}{\bibfnamefont{M.}~\bibnamefont{Kunz}}, \bibnamefont{and}
  \bibinfo{author}{\bibfnamefont{A.}~\bibnamefont{Melchiorri}},
  \bibinfo{journal}{Phys.Rept.} \textbf{\bibinfo{volume}{364}},
  \bibinfo{pages}{1} (\bibinfo{year}{2002}), \eprint{astro-ph/0110348}.

\bibitem[{\citenamefont{Copeland and Kibble}(2010)}]{Copeland:2009ga}
\bibinfo{author}{\bibfnamefont{E.~J.} \bibnamefont{Copeland}} \bibnamefont{and}
  \bibinfo{author}{\bibfnamefont{T.}~\bibnamefont{Kibble}},
  \bibinfo{journal}{Proc.Roy.Soc.Lond.} \textbf{\bibinfo{volume}{A466}},
  \bibinfo{pages}{623} (\bibinfo{year}{2010}), \eprint{0911.1345}.

\bibitem[{\citenamefont{Hindmarsh}(2011)}]{Hindmarsh:2011qj}
\bibinfo{author}{\bibfnamefont{M.}~\bibnamefont{Hindmarsh}},
  \bibinfo{journal}{Prog.Theor.Phys.Suppl.} \textbf{\bibinfo{volume}{190}},
  \bibinfo{pages}{197} (\bibinfo{year}{2011}), \eprint{1106.0391}.

\bibitem[{\citenamefont{Figueroa et~al.}(2013)\citenamefont{Figueroa,
  Hindmarsh, and Urrestilla}}]{Figueroa:2012kw}
\bibinfo{author}{\bibfnamefont{D.~G.} \bibnamefont{Figueroa}},
  \bibinfo{author}{\bibfnamefont{M.}~\bibnamefont{Hindmarsh}},
  \bibnamefont{and}
  \bibinfo{author}{\bibfnamefont{J.}~\bibnamefont{Urrestilla}},
  \bibinfo{journal}{Phys.Rev.Lett.} \textbf{\bibinfo{volume}{110}},
  \bibinfo{pages}{101302} (\bibinfo{year}{2013}), \eprint{1212.5458}.

\bibitem[{\citenamefont{Wyman et~al.}(2005)\citenamefont{Wyman, Pogosian, and
  Wasserman}}]{Wyman:2005tu}
\bibinfo{author}{\bibfnamefont{M.}~\bibnamefont{Wyman}},
  \bibinfo{author}{\bibfnamefont{L.}~\bibnamefont{Pogosian}}, \bibnamefont{and}
  \bibinfo{author}{\bibfnamefont{I.}~\bibnamefont{Wasserman}},
  \bibinfo{journal}{Phys.Rev.} \textbf{\bibinfo{volume}{D72}},
  \bibinfo{pages}{023513} (\bibinfo{year}{2005}), \eprint{astro-ph/0503364}.

\bibitem[{\citenamefont{Bevis et~al.}(2008)\citenamefont{Bevis, Hindmarsh,
  Kunz, and Urrestilla}}]{Bevis:2007gh}
\bibinfo{author}{\bibfnamefont{N.}~\bibnamefont{Bevis}},
  \bibinfo{author}{\bibfnamefont{M.}~\bibnamefont{Hindmarsh}},
  \bibinfo{author}{\bibfnamefont{M.}~\bibnamefont{Kunz}}, \bibnamefont{and}
  \bibinfo{author}{\bibfnamefont{J.}~\bibnamefont{Urrestilla}},
  \bibinfo{journal}{Phys.Rev.Lett.} \textbf{\bibinfo{volume}{100}},
  \bibinfo{pages}{021301} (\bibinfo{year}{2008}), \eprint{astro-ph/0702223}.

\bibitem[{\citenamefont{Battye and Moss}(2010)}]{Battye:2010xz}
\bibinfo{author}{\bibfnamefont{R.}~\bibnamefont{Battye}} \bibnamefont{and}
  \bibinfo{author}{\bibfnamefont{A.}~\bibnamefont{Moss}},
  \bibinfo{journal}{Phys.Rev.} \textbf{\bibinfo{volume}{D82}},
  \bibinfo{pages}{023521} (\bibinfo{year}{2010}), \eprint{1005.0479}.

\bibitem[{\citenamefont{Dunkley et~al.}(2011)\citenamefont{Dunkley, Hlozek,
  Sievers, Acquaviva, Ade et~al.}}]{Dunkley:2010ge}
\bibinfo{author}{\bibfnamefont{J.}~\bibnamefont{Dunkley}},
  \bibinfo{author}{\bibfnamefont{R.}~\bibnamefont{Hlozek}},
  \bibinfo{author}{\bibfnamefont{J.}~\bibnamefont{Sievers}},
  \bibinfo{author}{\bibfnamefont{V.}~\bibnamefont{Acquaviva}},
  \bibinfo{author}{\bibfnamefont{P.}~\bibnamefont{Ade}}, \bibnamefont{et~al.},
  \bibinfo{journal}{Astrophys.J.} \textbf{\bibinfo{volume}{739}},
  \bibinfo{pages}{52} (\bibinfo{year}{2011}), \eprint{1009.0866}.

\bibitem[{\citenamefont{Urrestilla et~al.}(2011)\citenamefont{Urrestilla,
  Bevis, Hindmarsh, and Kunz}}]{Urrestilla:2011gr}
\bibinfo{author}{\bibfnamefont{J.}~\bibnamefont{Urrestilla}},
  \bibinfo{author}{\bibfnamefont{N.}~\bibnamefont{Bevis}},
  \bibinfo{author}{\bibfnamefont{M.}~\bibnamefont{Hindmarsh}},
  \bibnamefont{and} \bibinfo{author}{\bibfnamefont{M.}~\bibnamefont{Kunz}},
  \bibinfo{journal}{JCAP} \textbf{\bibinfo{volume}{1112}}, \bibinfo{pages}{021}
  (\bibinfo{year}{2011}), \eprint{1108.2730}.

\bibitem[{\citenamefont{Avgoustidis et~al.}(2011)\citenamefont{Avgoustidis,
  Copeland, Moss, Pogosian, Pourtsidou et~al.}}]{Avgoustidis:2011ax}
\bibinfo{author}{\bibfnamefont{A.}~\bibnamefont{Avgoustidis}},
  \bibinfo{author}{\bibfnamefont{E.}~\bibnamefont{Copeland}},
  \bibinfo{author}{\bibfnamefont{A.}~\bibnamefont{Moss}},
  \bibinfo{author}{\bibfnamefont{L.}~\bibnamefont{Pogosian}},
  \bibinfo{author}{\bibfnamefont{A.}~\bibnamefont{Pourtsidou}},
  \bibnamefont{et~al.}, \bibinfo{journal}{Phys.Rev.Lett.}
  \textbf{\bibinfo{volume}{107}}, \bibinfo{pages}{121301}
  (\bibinfo{year}{2011}), \eprint{1105.6198}.

\bibitem[{\citenamefont{Ade et~al.}(2014{\natexlab{d}})}]{Ade:2013xla}
\bibinfo{author}{\bibfnamefont{P.}~\bibnamefont{Ade}} \bibnamefont{et~al.}
  (\bibinfo{collaboration}{Planck Collaboration}),
  \bibinfo{journal}{Astron.Astrophys.} \textbf{\bibinfo{volume}{571}},
  \bibinfo{pages}{A25} (\bibinfo{year}{2014}{\natexlab{d}}),
  \eprint{1303.5085},
  \urlprefix\url{http://dx.doi.org/10.1051/0004-6361/201321621}.

\bibitem[{\citenamefont{Nielsen and Olesen}(1973)}]{Nielsen:1973cs}
\bibinfo{author}{\bibfnamefont{H.~B.} \bibnamefont{Nielsen}} \bibnamefont{and}
  \bibinfo{author}{\bibfnamefont{P.}~\bibnamefont{Olesen}},
  \bibinfo{journal}{Nucl.Phys.} \textbf{\bibinfo{volume}{B61}},
  \bibinfo{pages}{45} (\bibinfo{year}{1973}).

\bibitem[{\citenamefont{Turok and Spergel}(1990)}]{Turok:1990gw}
\bibinfo{author}{\bibfnamefont{N.}~\bibnamefont{Turok}} \bibnamefont{and}
  \bibinfo{author}{\bibfnamefont{D.}~\bibnamefont{Spergel}},
  \bibinfo{journal}{Phys.Rev.Lett.} \textbf{\bibinfo{volume}{64}},
  \bibinfo{pages}{2736} (\bibinfo{year}{1990}).

\bibitem[{\citenamefont{Vachaspati and
  Ach{\'u}carro}(1991)}]{Vachaspati:1991dz}
\bibinfo{author}{\bibfnamefont{T.}~\bibnamefont{Vachaspati}} \bibnamefont{and}
  \bibinfo{author}{\bibfnamefont{A.}~\bibnamefont{Ach{\'u}carro}},
  \bibinfo{journal}{Phys.Rev.} \textbf{\bibinfo{volume}{D44}},
  \bibinfo{pages}{3067} (\bibinfo{year}{1991}).

\bibitem[{\citenamefont{Hindmarsh}(1992)}]{Hindmarsh:1991jq}
\bibinfo{author}{\bibfnamefont{M.}~\bibnamefont{Hindmarsh}},
  \bibinfo{journal}{Phys.Rev.Lett.} \textbf{\bibinfo{volume}{68}},
  \bibinfo{pages}{1263} (\bibinfo{year}{1992}).

\bibitem[{\citenamefont{Ach{\'u}carro and Vachaspati}(2000)}]{Achucarro:1999it}
\bibinfo{author}{\bibfnamefont{A.}~\bibnamefont{Ach{\'u}carro}}
  \bibnamefont{and}
  \bibinfo{author}{\bibfnamefont{T.}~\bibnamefont{Vachaspati}},
  \bibinfo{journal}{Phys.Rept.} \textbf{\bibinfo{volume}{327}},
  \bibinfo{pages}{347} (\bibinfo{year}{2000}), \eprint{hep-ph/9904229}.

\bibitem[{\citenamefont{Hindmarsh}(1993)}]{Hindmarsh:1992yy}
\bibinfo{author}{\bibfnamefont{M.}~\bibnamefont{Hindmarsh}},
  \bibinfo{journal}{Nucl.Phys.} \textbf{\bibinfo{volume}{B392}},
  \bibinfo{pages}{461} (\bibinfo{year}{1993}), \eprint{hep-ph/9206229}.

\bibitem[{\citenamefont{Bevis et~al.}(2007{\natexlab{a}})\citenamefont{Bevis,
  Hindmarsh, Kunz, and Urrestilla}}]{Bevis:2006mj}
\bibinfo{author}{\bibfnamefont{N.}~\bibnamefont{Bevis}},
  \bibinfo{author}{\bibfnamefont{M.}~\bibnamefont{Hindmarsh}},
  \bibinfo{author}{\bibfnamefont{M.}~\bibnamefont{Kunz}}, \bibnamefont{and}
  \bibinfo{author}{\bibfnamefont{J.}~\bibnamefont{Urrestilla}},
  \bibinfo{journal}{Phys.Rev.} \textbf{\bibinfo{volume}{D75}},
  \bibinfo{pages}{065015} (\bibinfo{year}{2007}{\natexlab{a}}),
  \eprint{astro-ph/0605018}.

\bibitem[{\citenamefont{Bevis et~al.}(2007{\natexlab{b}})\citenamefont{Bevis,
  Hindmarsh, Kunz, and Urrestilla}}]{Bevis:2007qz}
\bibinfo{author}{\bibfnamefont{N.}~\bibnamefont{Bevis}},
  \bibinfo{author}{\bibfnamefont{M.}~\bibnamefont{Hindmarsh}},
  \bibinfo{author}{\bibfnamefont{M.}~\bibnamefont{Kunz}}, \bibnamefont{and}
  \bibinfo{author}{\bibfnamefont{J.}~\bibnamefont{Urrestilla}},
  \bibinfo{journal}{Phys.Rev.} \textbf{\bibinfo{volume}{D76}},
  \bibinfo{pages}{043005} (\bibinfo{year}{2007}{\natexlab{b}}),
  \eprint{0704.3800}.

\bibitem[{\citenamefont{Bevis et~al.}(2010)\citenamefont{Bevis, Hindmarsh,
  Kunz, and Urrestilla}}]{Bevis:2010gj}
\bibinfo{author}{\bibfnamefont{N.}~\bibnamefont{Bevis}},
  \bibinfo{author}{\bibfnamefont{M.}~\bibnamefont{Hindmarsh}},
  \bibinfo{author}{\bibfnamefont{M.}~\bibnamefont{Kunz}}, \bibnamefont{and}
  \bibinfo{author}{\bibfnamefont{J.}~\bibnamefont{Urrestilla}},
  \bibinfo{journal}{Phys.Rev.} \textbf{\bibinfo{volume}{D82}},
  \bibinfo{pages}{065004} (\bibinfo{year}{2010}), \eprint{1005.2663}.

\bibitem[{\citenamefont{Pen et~al.}(1997)\citenamefont{Pen, Seljak, and
  Turok}}]{Pen:1997ae}
\bibinfo{author}{\bibfnamefont{U.-L.} \bibnamefont{Pen}},
  \bibinfo{author}{\bibfnamefont{U.}~\bibnamefont{Seljak}}, \bibnamefont{and}
  \bibinfo{author}{\bibfnamefont{N.}~\bibnamefont{Turok}},
  \bibinfo{journal}{Phys.Rev.Lett.} \textbf{\bibinfo{volume}{79}},
  \bibinfo{pages}{1611} (\bibinfo{year}{1997}), \eprint{astro-ph/9704165}.

\bibitem[{\citenamefont{Durrer et~al.}(1999)\citenamefont{Durrer, Kunz, and
  Melchiorri}}]{Durrer:1998rw}
\bibinfo{author}{\bibfnamefont{R.}~\bibnamefont{Durrer}},
  \bibinfo{author}{\bibfnamefont{M.}~\bibnamefont{Kunz}}, \bibnamefont{and}
  \bibinfo{author}{\bibfnamefont{A.}~\bibnamefont{Melchiorri}},
  \bibinfo{journal}{Phys.Rev.} \textbf{\bibinfo{volume}{D59}},
  \bibinfo{pages}{123005} (\bibinfo{year}{1999}), \eprint{astro-ph/9811174}.

\bibitem[{\citenamefont{Urrestilla
  et~al.}(2008{\natexlab{a}})\citenamefont{Urrestilla, Bevis, Hindmarsh, Kunz,
  and Liddle}}]{Urrestilla:2007sf}
\bibinfo{author}{\bibfnamefont{J.}~\bibnamefont{Urrestilla}},
  \bibinfo{author}{\bibfnamefont{N.}~\bibnamefont{Bevis}},
  \bibinfo{author}{\bibfnamefont{M.}~\bibnamefont{Hindmarsh}},
  \bibinfo{author}{\bibfnamefont{M.}~\bibnamefont{Kunz}}, \bibnamefont{and}
  \bibinfo{author}{\bibfnamefont{A.~R.} \bibnamefont{Liddle}},
  \bibinfo{journal}{JCAP} \textbf{\bibinfo{volume}{0807}}, \bibinfo{pages}{010}
  (\bibinfo{year}{2008}{\natexlab{a}}), \eprint{0711.1842}.

\bibitem[{\citenamefont{Urrestilla
  et~al.}(2008{\natexlab{b}})\citenamefont{Urrestilla, Mukherjee, Liddle,
  Bevis, Hindmarsh et~al.}}]{Urrestilla:2008jv}
\bibinfo{author}{\bibfnamefont{J.}~\bibnamefont{Urrestilla}},
  \bibinfo{author}{\bibfnamefont{P.}~\bibnamefont{Mukherjee}},
  \bibinfo{author}{\bibfnamefont{A.~R.} \bibnamefont{Liddle}},
  \bibinfo{author}{\bibfnamefont{N.}~\bibnamefont{Bevis}},
  \bibinfo{author}{\bibfnamefont{M.}~\bibnamefont{Hindmarsh}},
  \bibnamefont{et~al.}, \bibinfo{journal}{Phys.Rev.}
  \textbf{\bibinfo{volume}{D77}}, \bibinfo{pages}{123005}
  (\bibinfo{year}{2008}{\natexlab{b}}), \eprint{0803.2059}.

\bibitem[{\citenamefont{Copeland et~al.}(2011)\citenamefont{Copeland, Pogosian,
  and Vachaspati}}]{Copeland:2011dx}
\bibinfo{author}{\bibfnamefont{E.~J.} \bibnamefont{Copeland}},
  \bibinfo{author}{\bibfnamefont{L.}~\bibnamefont{Pogosian}}, \bibnamefont{and}
  \bibinfo{author}{\bibfnamefont{T.}~\bibnamefont{Vachaspati}},
  \bibinfo{journal}{Class.Quant.Grav.} \textbf{\bibinfo{volume}{28}},
  \bibinfo{pages}{204009} (\bibinfo{year}{2011}), \eprint{1105.0207}.

\bibitem[{\citenamefont{Yokoyama}(1988)}]{Yokoyama:1988zza}
\bibinfo{author}{\bibfnamefont{J.}~\bibnamefont{Yokoyama}},
  \bibinfo{journal}{Phys.Lett.} \textbf{\bibinfo{volume}{B212}},
  \bibinfo{pages}{273} (\bibinfo{year}{1988}).

\bibitem[{\citenamefont{Copeland et~al.}(1994)\citenamefont{Copeland, Liddle,
  Lyth, Stewart, and Wands}}]{Copeland:1994vg}
\bibinfo{author}{\bibfnamefont{E.~J.} \bibnamefont{Copeland}},
  \bibinfo{author}{\bibfnamefont{A.~R.} \bibnamefont{Liddle}},
  \bibinfo{author}{\bibfnamefont{D.~H.} \bibnamefont{Lyth}},
  \bibinfo{author}{\bibfnamefont{E.~D.} \bibnamefont{Stewart}},
  \bibnamefont{and} \bibinfo{author}{\bibfnamefont{D.}~\bibnamefont{Wands}},
  \bibinfo{journal}{Phys.Rev.} \textbf{\bibinfo{volume}{D49}},
  \bibinfo{pages}{6410} (\bibinfo{year}{1994}), \eprint{astro-ph/9401011}.

\bibitem[{\citenamefont{Lyth and Riotto}(1999)}]{Lyth:1998xn}
\bibinfo{author}{\bibfnamefont{D.~H.} \bibnamefont{Lyth}} \bibnamefont{and}
  \bibinfo{author}{\bibfnamefont{A.}~\bibnamefont{Riotto}},
  \bibinfo{journal}{Phys.Rept.} \textbf{\bibinfo{volume}{314}},
  \bibinfo{pages}{1} (\bibinfo{year}{1999}), \eprint{hep-ph/9807278}.

\bibitem[{\citenamefont{Jeannerot et~al.}(2003)\citenamefont{Jeannerot, Rocher,
  and Sakellariadou}}]{Jeannerot:2003qv}
\bibinfo{author}{\bibfnamefont{R.}~\bibnamefont{Jeannerot}},
  \bibinfo{author}{\bibfnamefont{J.}~\bibnamefont{Rocher}}, \bibnamefont{and}
  \bibinfo{author}{\bibfnamefont{M.}~\bibnamefont{Sakellariadou}},
  \bibinfo{journal}{Phys.Rev.} \textbf{\bibinfo{volume}{D68}},
  \bibinfo{pages}{103514} (\bibinfo{year}{2003}), \eprint{hep-ph/0308134}.

\bibitem[{\citenamefont{Majumdar and Christine-Davis}(2002)}]{Majumdar:2002hy}
\bibinfo{author}{\bibfnamefont{M.}~\bibnamefont{Majumdar}} \bibnamefont{and}
  \bibinfo{author}{\bibfnamefont{A.}~\bibnamefont{Christine-Davis}},
  \bibinfo{journal}{JHEP} \textbf{\bibinfo{volume}{0203}}, \bibinfo{pages}{056}
  (\bibinfo{year}{2002}), \eprint{hep-th/0202148}.

\bibitem[{\citenamefont{Sarangi and Tye}(2002)}]{Sarangi:2002yt}
\bibinfo{author}{\bibfnamefont{S.}~\bibnamefont{Sarangi}} \bibnamefont{and}
  \bibinfo{author}{\bibfnamefont{S.~H.} \bibnamefont{Tye}},
  \bibinfo{journal}{Phys.Lett.} \textbf{\bibinfo{volume}{B536}},
  \bibinfo{pages}{185} (\bibinfo{year}{2002}), \eprint{hep-th/0204074}.

\bibitem[{\citenamefont{Copeland et~al.}(2004)\citenamefont{Copeland, Myers,
  and Polchinski}}]{Copeland:2003bj}
\bibinfo{author}{\bibfnamefont{E.~J.} \bibnamefont{Copeland}},
  \bibinfo{author}{\bibfnamefont{R.~C.} \bibnamefont{Myers}}, \bibnamefont{and}
  \bibinfo{author}{\bibfnamefont{J.}~\bibnamefont{Polchinski}},
  \bibinfo{journal}{JHEP} \textbf{\bibinfo{volume}{0406}}, \bibinfo{pages}{013}
  (\bibinfo{year}{2004}), \eprint{hep-th/0312067}.

\bibitem[{\citenamefont{Dvali and Vilenkin}(2004)}]{Dvali:2003zj}
\bibinfo{author}{\bibfnamefont{G.}~\bibnamefont{Dvali}} \bibnamefont{and}
  \bibinfo{author}{\bibfnamefont{A.}~\bibnamefont{Vilenkin}},
  \bibinfo{journal}{JCAP} \textbf{\bibinfo{volume}{0403}}, \bibinfo{pages}{010}
  (\bibinfo{year}{2004}), \eprint{hep-th/0312007}.

\bibitem[{\citenamefont{Tye and Wong}(2014)}]{Tye:2014tja}
\bibinfo{author}{\bibfnamefont{S.~H.~H.} \bibnamefont{Tye}} \bibnamefont{and}
  \bibinfo{author}{\bibfnamefont{S.~S.~C.} \bibnamefont{Wong}}
  (\bibinfo{year}{2014}), \eprint{1404.6988}.

\bibitem[{\citenamefont{Albrecht and Turok}(1985)}]{Albrecht:1984xv}
\bibinfo{author}{\bibfnamefont{A.}~\bibnamefont{Albrecht}} \bibnamefont{and}
  \bibinfo{author}{\bibfnamefont{N.}~\bibnamefont{Turok}},
  \bibinfo{journal}{Phys.Rev.Lett.} \textbf{\bibinfo{volume}{54}},
  \bibinfo{pages}{1868} (\bibinfo{year}{1985}).

\bibitem[{\citenamefont{Bennett and Bouchet}(1988)}]{Bennett:1987vf}
\bibinfo{author}{\bibfnamefont{D.~P.} \bibnamefont{Bennett}} \bibnamefont{and}
  \bibinfo{author}{\bibfnamefont{F.~R.} \bibnamefont{Bouchet}},
  \bibinfo{journal}{Phys.Rev.Lett.} \textbf{\bibinfo{volume}{60}},
  \bibinfo{pages}{257} (\bibinfo{year}{1988}).

\bibitem[{\citenamefont{Albrecht and Turok}(1989)}]{Albrecht:1989mk}
\bibinfo{author}{\bibfnamefont{A.}~\bibnamefont{Albrecht}} \bibnamefont{and}
  \bibinfo{author}{\bibfnamefont{N.}~\bibnamefont{Turok}},
  \bibinfo{journal}{Phys.Rev.} \textbf{\bibinfo{volume}{D40}},
  \bibinfo{pages}{973} (\bibinfo{year}{1989}).

\bibitem[{\citenamefont{Bennett and Bouchet}(1989)}]{Bennett:1989ak}
\bibinfo{author}{\bibfnamefont{D.~P.} \bibnamefont{Bennett}} \bibnamefont{and}
  \bibinfo{author}{\bibfnamefont{F.~R.} \bibnamefont{Bouchet}},
  \bibinfo{journal}{Phys.Rev.Lett.} \textbf{\bibinfo{volume}{63}},
  \bibinfo{pages}{2776} (\bibinfo{year}{1989}).

\bibitem[{\citenamefont{Bennett and Bouchet}(1990)}]{Bennett:1989yp}
\bibinfo{author}{\bibfnamefont{D.~P.} \bibnamefont{Bennett}} \bibnamefont{and}
  \bibinfo{author}{\bibfnamefont{F.~R.} \bibnamefont{Bouchet}},
  \bibinfo{journal}{Phys.Rev.} \textbf{\bibinfo{volume}{D41}},
  \bibinfo{pages}{2408} (\bibinfo{year}{1990}).

\bibitem[{\citenamefont{Allen and Shellard}(1990)}]{Allen:1990tv}
\bibinfo{author}{\bibfnamefont{B.}~\bibnamefont{Allen}} \bibnamefont{and}
  \bibinfo{author}{\bibfnamefont{E.}~\bibnamefont{Shellard}},
  \bibinfo{journal}{Phys.Rev.Lett.} \textbf{\bibinfo{volume}{64}},
  \bibinfo{pages}{119} (\bibinfo{year}{1990}).

\bibitem[{\citenamefont{Sakellariadou and
  Vilenkin}(1990)}]{Sakellariadou:1990nd}
\bibinfo{author}{\bibfnamefont{M.}~\bibnamefont{Sakellariadou}}
  \bibnamefont{and} \bibinfo{author}{\bibfnamefont{A.}~\bibnamefont{Vilenkin}},
  \bibinfo{journal}{Phys.Rev.} \textbf{\bibinfo{volume}{D42}},
  \bibinfo{pages}{349} (\bibinfo{year}{1990}).

\bibitem[{\citenamefont{Vincent et~al.}(1997)\citenamefont{Vincent, Hindmarsh,
  and Sakellariadou}}]{Vincent:1996rb}
\bibinfo{author}{\bibfnamefont{G.~R.} \bibnamefont{Vincent}},
  \bibinfo{author}{\bibfnamefont{M.}~\bibnamefont{Hindmarsh}},
  \bibnamefont{and}
  \bibinfo{author}{\bibfnamefont{M.}~\bibnamefont{Sakellariadou}},
  \bibinfo{journal}{Phys.Rev.} \textbf{\bibinfo{volume}{D56}},
  \bibinfo{pages}{637} (\bibinfo{year}{1997}), \eprint{astro-ph/9612135}.

\bibitem[{\citenamefont{Vanchurin et~al.}(2005)\citenamefont{Vanchurin, Olum,
  and Vilenkin}}]{Vanchurin:2005yb}
\bibinfo{author}{\bibfnamefont{V.}~\bibnamefont{Vanchurin}},
  \bibinfo{author}{\bibfnamefont{K.}~\bibnamefont{Olum}}, \bibnamefont{and}
  \bibinfo{author}{\bibfnamefont{A.}~\bibnamefont{Vilenkin}},
  \bibinfo{journal}{Phys.Rev.} \textbf{\bibinfo{volume}{D72}},
  \bibinfo{pages}{063514} (\bibinfo{year}{2005}), \eprint{gr-qc/0501040}.

\bibitem[{\citenamefont{Vanchurin et~al.}(2006)\citenamefont{Vanchurin, Olum,
  and Vilenkin}}]{Vanchurin:2005pa}
\bibinfo{author}{\bibfnamefont{V.}~\bibnamefont{Vanchurin}},
  \bibinfo{author}{\bibfnamefont{K.~D.} \bibnamefont{Olum}}, \bibnamefont{and}
  \bibinfo{author}{\bibfnamefont{A.}~\bibnamefont{Vilenkin}},
  \bibinfo{journal}{Phys.Rev.} \textbf{\bibinfo{volume}{D74}},
  \bibinfo{pages}{063527} (\bibinfo{year}{2006}), \eprint{gr-qc/0511159}.

\bibitem[{\citenamefont{Ringeval et~al.}(2007)\citenamefont{Ringeval,
  Sakellariadou, and Bouchet}}]{Ringeval:2005kr}
\bibinfo{author}{\bibfnamefont{C.}~\bibnamefont{Ringeval}},
  \bibinfo{author}{\bibfnamefont{M.}~\bibnamefont{Sakellariadou}},
  \bibnamefont{and} \bibinfo{author}{\bibfnamefont{F.}~\bibnamefont{Bouchet}},
  \bibinfo{journal}{JCAP} \textbf{\bibinfo{volume}{0702}}, \bibinfo{pages}{023}
  (\bibinfo{year}{2007}), \eprint{astro-ph/0511646}.

\bibitem[{\citenamefont{Olum and Vanchurin}(2007)}]{Olum:2006ix}
\bibinfo{author}{\bibfnamefont{K.~D.} \bibnamefont{Olum}} \bibnamefont{and}
  \bibinfo{author}{\bibfnamefont{V.}~\bibnamefont{Vanchurin}},
  \bibinfo{journal}{Phys.Rev.} \textbf{\bibinfo{volume}{D75}},
  \bibinfo{pages}{063521} (\bibinfo{year}{2007}), \eprint{astro-ph/0610419}.

\bibitem[{\citenamefont{Lorenz et~al.}(2010)\citenamefont{Lorenz, Ringeval, and
  Sakellariadou}}]{Lorenz:2010sm}
\bibinfo{author}{\bibfnamefont{L.}~\bibnamefont{Lorenz}},
  \bibinfo{author}{\bibfnamefont{C.}~\bibnamefont{Ringeval}}, \bibnamefont{and}
  \bibinfo{author}{\bibfnamefont{M.}~\bibnamefont{Sakellariadou}},
  \bibinfo{journal}{JCAP} \textbf{\bibinfo{volume}{1010}}, \bibinfo{pages}{003}
  (\bibinfo{year}{2010}), \eprint{1006.0931}.

\bibitem[{\citenamefont{Blanco-Pillado
  et~al.}(2012)\citenamefont{Blanco-Pillado, Olum, and
  Shlaer}}]{BlancoPillado:2010sy}
\bibinfo{author}{\bibfnamefont{J.~J.} \bibnamefont{Blanco-Pillado}},
  \bibinfo{author}{\bibfnamefont{K.~D.} \bibnamefont{Olum}}, \bibnamefont{and}
  \bibinfo{author}{\bibfnamefont{B.}~\bibnamefont{Shlaer}},
  \bibinfo{journal}{J.Comput.Phys.} \textbf{\bibinfo{volume}{231}},
  \bibinfo{pages}{98} (\bibinfo{year}{2012}), \eprint{1011.4046}.

\bibitem[{\citenamefont{Blanco-Pillado
  et~al.}(2011)\citenamefont{Blanco-Pillado, Olum, and
  Shlaer}}]{BlancoPillado:2011dq}
\bibinfo{author}{\bibfnamefont{J.~J.} \bibnamefont{Blanco-Pillado}},
  \bibinfo{author}{\bibfnamefont{K.~D.} \bibnamefont{Olum}}, \bibnamefont{and}
  \bibinfo{author}{\bibfnamefont{B.}~\bibnamefont{Shlaer}},
  \bibinfo{journal}{Phys.Rev.} \textbf{\bibinfo{volume}{D83}},
  \bibinfo{pages}{083514} (\bibinfo{year}{2011}), \eprint{1101.5173}.

\bibitem[{\citenamefont{Landriau and Shellard}(2011)}]{Landriau:2010cb}
\bibinfo{author}{\bibfnamefont{M.}~\bibnamefont{Landriau}} \bibnamefont{and}
  \bibinfo{author}{\bibfnamefont{E.}~\bibnamefont{Shellard}},
  \bibinfo{journal}{Phys.Rev.} \textbf{\bibinfo{volume}{D83}},
  \bibinfo{pages}{043516} (\bibinfo{year}{2011}), \eprint{1004.2885}.

\bibitem[{\citenamefont{Ringeval}(2010)}]{Ringeval:2010ca}
\bibinfo{author}{\bibfnamefont{C.}~\bibnamefont{Ringeval}},
  \bibinfo{journal}{Adv.Astron.} \textbf{\bibinfo{volume}{2010}},
  \bibinfo{pages}{380507} (\bibinfo{year}{2010}), \eprint{1005.4842}.

\bibitem[{\citenamefont{Albrecht et~al.}(1998)\citenamefont{Albrecht, Battye,
  and Robinson}}]{Albrecht:1997mz}
\bibinfo{author}{\bibfnamefont{A.}~\bibnamefont{Albrecht}},
  \bibinfo{author}{\bibfnamefont{R.~A.} \bibnamefont{Battye}},
  \bibnamefont{and} \bibinfo{author}{\bibfnamefont{J.}~\bibnamefont{Robinson}},
  \bibinfo{journal}{Phys.Rev.} \textbf{\bibinfo{volume}{D59}},
  \bibinfo{pages}{023508} (\bibinfo{year}{1998}), \eprint{astro-ph/9711121}.

\bibitem[{\citenamefont{Avgoustidis et~al.}(2012)\citenamefont{Avgoustidis,
  Copeland, Moss, and Skliros}}]{Avgoustidis:2012gb}
\bibinfo{author}{\bibfnamefont{A.}~\bibnamefont{Avgoustidis}},
  \bibinfo{author}{\bibfnamefont{E.~J.} \bibnamefont{Copeland}},
  \bibinfo{author}{\bibfnamefont{A.}~\bibnamefont{Moss}}, \bibnamefont{and}
  \bibinfo{author}{\bibfnamefont{D.}~\bibnamefont{Skliros}},
  \bibinfo{journal}{Phys.Rev.} \textbf{\bibinfo{volume}{D86}},
  \bibinfo{pages}{123513} (\bibinfo{year}{2012}), \eprint{1209.2461}.

\bibitem[{\citenamefont{Pogosian and Vachaspati}(1999)}]{Pogosian:1999np}
\bibinfo{author}{\bibfnamefont{L.}~\bibnamefont{Pogosian}} \bibnamefont{and}
  \bibinfo{author}{\bibfnamefont{T.}~\bibnamefont{Vachaspati}},
  \bibinfo{journal}{Phys.Rev.} \textbf{\bibinfo{volume}{D60}},
  \bibinfo{pages}{083504} (\bibinfo{year}{1999}), \eprint{astro-ph/9903361}.

\bibitem[{\citenamefont{Kunz and Durrer}(1997)}]{Kunz:1996ka}
\bibinfo{author}{\bibfnamefont{M.}~\bibnamefont{Kunz}} \bibnamefont{and}
  \bibinfo{author}{\bibfnamefont{R.}~\bibnamefont{Durrer}},
  \bibinfo{journal}{Phys.Rev.} \textbf{\bibinfo{volume}{D55}},
  \bibinfo{pages}{R4516} (\bibinfo{year}{1997}), \eprint{astro-ph/9612202}.

\bibitem[{\citenamefont{Nunes et~al.}(2011)\citenamefont{Nunes, Avgoustidis,
  Martins, and Urrestilla}}]{Nunes:2011sf}
\bibinfo{author}{\bibfnamefont{A.}~\bibnamefont{Nunes}},
  \bibinfo{author}{\bibfnamefont{A.}~\bibnamefont{Avgoustidis}},
  \bibinfo{author}{\bibfnamefont{C.}~\bibnamefont{Martins}}, \bibnamefont{and}
  \bibinfo{author}{\bibfnamefont{J.}~\bibnamefont{Urrestilla}},
  \bibinfo{journal}{Phys.Rev.} \textbf{\bibinfo{volume}{D84}},
  \bibinfo{pages}{063504} (\bibinfo{year}{2011}), \eprint{1107.2008}.

\bibitem[{\citenamefont{Ach{\'u}carro et~al.}(2014)\citenamefont{Ach{\'u}carro,
  Avgoustidis, Leite, Lopez-Eiguren, Martins et~al.}}]{Achucarro:2013mga}
\bibinfo{author}{\bibfnamefont{A.}~\bibnamefont{Ach{\'u}carro}},
  \bibinfo{author}{\bibfnamefont{A.}~\bibnamefont{Avgoustidis}},
  \bibinfo{author}{\bibfnamefont{A.}~\bibnamefont{Leite}},
  \bibinfo{author}{\bibfnamefont{A.}~\bibnamefont{Lopez-Eiguren}},
  \bibinfo{author}{\bibfnamefont{C.}~\bibnamefont{Martins}},
  \bibnamefont{et~al.}, \bibinfo{journal}{Phys.Rev.}
  \textbf{\bibinfo{volume}{D89}}, \bibinfo{pages}{063503}
  (\bibinfo{year}{2014}), \eprint{1312.2123}.

\bibitem[{\citenamefont{Doran}(2005)}]{Doran:2003sy}
\bibinfo{author}{\bibfnamefont{M.}~\bibnamefont{Doran}},
  \bibinfo{journal}{JCAP} \textbf{\bibinfo{volume}{0510}}, \bibinfo{pages}{011}
  (\bibinfo{year}{2005}), \eprint{astro-ph/0302138}.

\bibitem[{\citenamefont{Komatsu et~al.}(2011)}]{Komatsu:2010fb}
\bibinfo{author}{\bibfnamefont{E.}~\bibnamefont{Komatsu}} \bibnamefont{et~al.}
  (\bibinfo{collaboration}{WMAP Collaboration}),
  \bibinfo{journal}{Astrophys.J.Suppl.} \textbf{\bibinfo{volume}{192}},
  \bibinfo{pages}{18} (\bibinfo{year}{2011}), \eprint{1001.4538}.

\bibitem[{\citenamefont{Liddle and Lyth}(1992)}]{liddle:1992wi}
\bibinfo{author}{\bibfnamefont{A.~R.} \bibnamefont{Liddle}} \bibnamefont{and}
  \bibinfo{author}{\bibfnamefont{D.~H.} \bibnamefont{Lyth}},
  \bibinfo{journal}{Phys.Lett.} \textbf{\bibinfo{volume}{B291}},
  \bibinfo{pages}{391} (\bibinfo{year}{1992}), \eprint{astro-ph/9208007}.

\bibitem[{\citenamefont{Kamada et~al.}(2014)\citenamefont{Kamada, Miyamoto,
  Yamauchi, and Yokoyama}}]{Kamada:2014qta}
\bibinfo{author}{\bibfnamefont{K.}~\bibnamefont{Kamada}},
  \bibinfo{author}{\bibfnamefont{Y.}~\bibnamefont{Miyamoto}},
  \bibinfo{author}{\bibfnamefont{D.}~\bibnamefont{Yamauchi}}, \bibnamefont{and}
  \bibinfo{author}{\bibfnamefont{J.}~\bibnamefont{Yokoyama}},
  \bibinfo{journal}{Phys.Rev.} \textbf{\bibinfo{volume}{D90}},
  \bibinfo{pages}{083502} (\bibinfo{year}{2014}), \eprint{1407.2951}.

\bibitem[{\citenamefont{Audren et~al.}(2013)\citenamefont{Audren, Lesgourgues,
  Benabed, and Prunet}}]{Audren:2012wb}
\bibinfo{author}{\bibfnamefont{B.}~\bibnamefont{Audren}},
  \bibinfo{author}{\bibfnamefont{J.}~\bibnamefont{Lesgourgues}},
  \bibinfo{author}{\bibfnamefont{K.}~\bibnamefont{Benabed}}, \bibnamefont{and}
  \bibinfo{author}{\bibfnamefont{S.}~\bibnamefont{Prunet}},
  \bibinfo{journal}{JCAP} \textbf{\bibinfo{volume}{1302}}, \bibinfo{pages}{001}
  (\bibinfo{year}{2013}), \eprint{1210.7183}.

\bibitem[{Mon()}]{MontePython}
\bibinfo{howpublished}{See \url{http://montepython.net}}.

\bibitem[{\citenamefont{Lesgourgues}(2011)}]{Lesgourgues:2011re}
\bibinfo{author}{\bibfnamefont{J.}~\bibnamefont{Lesgourgues}}
  (\bibinfo{year}{2011}), \eprint{1104.2932}.

\bibitem[{\citenamefont{Blas et~al.}(2011)\citenamefont{Blas, Lesgourgues, and
  Tram}}]{Blas:2011rf}
\bibinfo{author}{\bibfnamefont{D.}~\bibnamefont{Blas}},
  \bibinfo{author}{\bibfnamefont{J.}~\bibnamefont{Lesgourgues}},
  \bibnamefont{and} \bibinfo{author}{\bibfnamefont{T.}~\bibnamefont{Tram}},
  \bibinfo{journal}{JCAP} \textbf{\bibinfo{volume}{1107}}, \bibinfo{pages}{034}
  (\bibinfo{year}{2011}), \eprint{1104.2933}.

\bibitem[{\citenamefont{Ade et~al.}(2014{\natexlab{e}})}]{Ade:2013zuv}
\bibinfo{author}{\bibfnamefont{P.}~\bibnamefont{Ade}} \bibnamefont{et~al.}
  (\bibinfo{collaboration}{Planck Collaboration}),
  \bibinfo{journal}{Astron.Astrophys.} \textbf{\bibinfo{volume}{571}},
  \bibinfo{pages}{A16} (\bibinfo{year}{2014}{\natexlab{e}}),
  \eprint{1303.5076}.

\bibitem[{\citenamefont{Hinshaw et~al.}(2013)}]{Hinshaw:2012aka}
\bibinfo{author}{\bibfnamefont{G.}~\bibnamefont{Hinshaw}} \bibnamefont{et~al.}
  (\bibinfo{collaboration}{WMAP}), \bibinfo{journal}{Astrophys.J.Suppl.}
  \textbf{\bibinfo{volume}{208}}, \bibinfo{pages}{19} (\bibinfo{year}{2013}),
  \eprint{1212.5226}.

\bibitem[{\citenamefont{Story et~al.}(2013)\citenamefont{Story, Reichardt, Hou,
  Keisler, Aird et~al.}}]{Story:2012wx}
\bibinfo{author}{\bibfnamefont{K.}~\bibnamefont{Story}},
  \bibinfo{author}{\bibfnamefont{C.}~\bibnamefont{Reichardt}},
  \bibinfo{author}{\bibfnamefont{Z.}~\bibnamefont{Hou}},
  \bibinfo{author}{\bibfnamefont{R.}~\bibnamefont{Keisler}},
  \bibinfo{author}{\bibfnamefont{K.}~\bibnamefont{Aird}}, \bibnamefont{et~al.},
  \bibinfo{journal}{Astrophys.J.} \textbf{\bibinfo{volume}{779}},
  \bibinfo{pages}{86} (\bibinfo{year}{2013}), \eprint{1210.7231}.

\bibitem[{\citenamefont{Reichardt et~al.}(2012)\citenamefont{Reichardt, Shaw,
  Zahn, Aird, Benson et~al.}}]{Reichardt:2011yv}
\bibinfo{author}{\bibfnamefont{C.}~\bibnamefont{Reichardt}},
  \bibinfo{author}{\bibfnamefont{L.}~\bibnamefont{Shaw}},
  \bibinfo{author}{\bibfnamefont{O.}~\bibnamefont{Zahn}},
  \bibinfo{author}{\bibfnamefont{K.}~\bibnamefont{Aird}},
  \bibinfo{author}{\bibfnamefont{B.}~\bibnamefont{Benson}},
  \bibnamefont{et~al.}, \bibinfo{journal}{Astrophys.J.}
  \textbf{\bibinfo{volume}{755}}, \bibinfo{pages}{70} (\bibinfo{year}{2012}),
  \eprint{1111.0932}.

\bibitem[{\citenamefont{Sievers et~al.}(2013)}]{Sievers:2013ica}
\bibinfo{author}{\bibfnamefont{J.~L.} \bibnamefont{Sievers}}
  \bibnamefont{et~al.} (\bibinfo{collaboration}{Atacama Cosmology Telescope}),
  \bibinfo{journal}{JCAP} \textbf{\bibinfo{volume}{1310}}, \bibinfo{pages}{060}
  (\bibinfo{year}{2013}), \eprint{1301.0824}.

\bibitem[{\citenamefont{Audren et~al.}(2014)\citenamefont{Audren, Figueroa, and
  Tram}}]{Audren:2014cea}
\bibinfo{author}{\bibfnamefont{B.}~\bibnamefont{Audren}},
  \bibinfo{author}{\bibfnamefont{D.~G.} \bibnamefont{Figueroa}},
  \bibnamefont{and} \bibinfo{author}{\bibfnamefont{T.}~\bibnamefont{Tram}}
  (\bibinfo{year}{2014}), \eprint{1405.1390}.

\bibitem[{\citenamefont{Smith et~al.}(2014)\citenamefont{Smith, Dvorkin, Boyle,
  Turok, Halpern et~al.}}]{Smith:2014kka}
\bibinfo{author}{\bibfnamefont{K.~M.} \bibnamefont{Smith}},
  \bibinfo{author}{\bibfnamefont{C.}~\bibnamefont{Dvorkin}},
  \bibinfo{author}{\bibfnamefont{L.}~\bibnamefont{Boyle}},
  \bibinfo{author}{\bibfnamefont{N.}~\bibnamefont{Turok}},
  \bibinfo{author}{\bibfnamefont{M.}~\bibnamefont{Halpern}},
  \bibnamefont{et~al.}, \bibinfo{journal}{Phys.Rev.Lett.}
  \textbf{\bibinfo{volume}{113}}, \bibinfo{pages}{031301}
  (\bibinfo{year}{2014}), \eprint{1404.0373}.

\bibitem[{\citenamefont{Martin et~al.}(2014)\citenamefont{Martin, Ringeval,
  Trotta, and Vennin}}]{Martin:2014lra}
\bibinfo{author}{\bibfnamefont{J.}~\bibnamefont{Martin}},
  \bibinfo{author}{\bibfnamefont{C.}~\bibnamefont{Ringeval}},
  \bibinfo{author}{\bibfnamefont{R.}~\bibnamefont{Trotta}}, \bibnamefont{and}
  \bibinfo{author}{\bibfnamefont{V.}~\bibnamefont{Vennin}}
  (\bibinfo{year}{2014}), \eprint{1405.7272}.

\bibitem[{Pla()}]{PlanckDust}
\bibinfo{howpublished}{See
  \url{http://rssd.esa.int/SA/PLANCK/docs/eslab47/Session09_Data_Processing/47ESLAB_April_04_10_30_Aumont.pdf}}.

\bibitem[{\citenamefont{Ade et~al.}(2014{\natexlab{f}})}]{Ade:2013kta}
\bibinfo{author}{\bibfnamefont{P.}~\bibnamefont{Ade}} \bibnamefont{et~al.}
  (\bibinfo{collaboration}{Planck Collaboration}),
  \bibinfo{journal}{Astron.Astrophys.} \textbf{\bibinfo{volume}{571}},
  \bibinfo{pages}{A15} (\bibinfo{year}{2014}{\natexlab{f}}),
  \eprint{1303.5075}.

\bibitem[{\citenamefont{Mukherjee et~al.}(2011)\citenamefont{Mukherjee,
  Urrestilla, Kunz, Liddle, Bevis et~al.}}]{Mukherjee:2010ve}
\bibinfo{author}{\bibfnamefont{P.}~\bibnamefont{Mukherjee}},
  \bibinfo{author}{\bibfnamefont{J.}~\bibnamefont{Urrestilla}},
  \bibinfo{author}{\bibfnamefont{M.}~\bibnamefont{Kunz}},
  \bibinfo{author}{\bibfnamefont{A.~R.} \bibnamefont{Liddle}},
  \bibinfo{author}{\bibfnamefont{N.}~\bibnamefont{Bevis}},
  \bibnamefont{et~al.}, \bibinfo{journal}{Phys.Rev.}
  \textbf{\bibinfo{volume}{D83}}, \bibinfo{pages}{043003}
  (\bibinfo{year}{2011}), \eprint{1010.5662}.

\bibitem[{\citenamefont{Adam et~al.}(2014)}]{Adam:2014bub}
\bibinfo{author}{\bibfnamefont{R.}~\bibnamefont{Adam}} \bibnamefont{et~al.}
  (\bibinfo{collaboration}{Planck Collaboration}) (\bibinfo{year}{2014}),
  \eprint{1409.5738}.

\end{thebibliography}

\end{document}